\documentclass[floats,floatfix,twocolumn,nosuperscriptaddress,nofootinbib,showpacs]{revtex4-1}

\pdfoutput=1
\usepackage{amsfonts,amssymb,amsmath}
\usepackage[usenames,dvipsnames]{xcolor}
\usepackage[unicode,colorlinks=true,linkcolor=blue,urlcolor=blue,citecolor=gray]{hyperref}
\usepackage[all]{hypcap}
\usepackage{graphicx}
\usepackage{xspace}
\usepackage{tikz}
\usetikzlibrary{shapes,arrows}
\usetikzlibrary{matrix}
\usetikzlibrary{shapes.multipart}
\usetikzlibrary{er,positioning}
\usepackage[normalem]{ulem} 
\usepackage{url}
\usepackage{color,soul}
\usepackage[caption=false]{subfig}
\usepackage[T1]{fontenc}
\usepackage[utf8]{inputenc}
\usepackage{microtype}

\newcommand{\Caltech}{\affiliation{Theoretical Astrophysics,
    Walter Burke Institute for Theoretical Physics,\\
    California Institute of Technology, Pasadena, CA 91125, USA}}
\newcommand{\Cornell}{\affiliation{Center for Astrophysics and Planetary Science, Cornell University, Ithaca, New York 14853, USA}}

\newcommand{\eps}{\ensuremath{\varepsilon}}

\newcommand{\pd}{\partial}

\newcommand{\nn}{\nonumber}

\newcommand{\txt}[1]{{\textrm{\tiny{#1}}}}
\newcommand{\mpl}{\ensuremath{m_\txt{pl}}}

\newcommand{\dual}{\,{}^*\!}

\begin{document}

\title{Evolving Metric Perturbations in dynamical Chern-Simons Gravity}

\author{Maria Okounkova}
\email{mokounko@tapir.caltech.edu}
\Caltech
\author{Mark A. Scheel}
\Caltech
\author{Saul A. Teukolsky}
\Caltech \Cornell

\date{\today}

\begin{abstract}
The stability of rotating black holes in dynamical Chern-Simons gravity (dCS) is an open question. To study this issue, we evolve the leading-order metric perturbation in order-reduced dynamical Chern-Simons gravity. The source is the leading-order dCS scalar field coupled to the spacetime curvature of a rotating black hole background. We use a well-posed, constraint-preserving scheme. We find that the leading-order metric perturbation numerically exhibits linear growth, but that the level of this growth converges to zero with numerical resolution. This analysis shows that spinning black holes in dCS gravity are numerically
stable to leading-order perturbations in the metric.
\end{abstract}


\maketitle

\section{Introduction}

Einstein's theory of general relativity (GR) has passed all precision tests to date and binary black hole observations from the Laser Interferometry Gravitational Wave Observatory (LIGO) have given a roughly 96\% agreement with GR~\cite{Will:2014kxa, TheLIGOScientific:2016src}. At some scale, however, GR must be reconciled with quantum mechanics in a quantum theory of gravity. Black hole systems can potentially illuminate signatures of quantum gravity, as they probe the strong-field, non-linear, high-curvature regime of gravity. 

While several null-hypothesis and parametrized tests of GR with LIGO observations have been performed~\cite{TheLIGOScientific:2016src, Yunes:2016jcc}, an open problem is the simulation of binary black holes through full inspiral, merger, and ringdown in a beyond-GR theory. Waveform predictions from such simulations would allow us to perform \textit{model-dependent} tests, and to parametrize the behavior at merger in beyond-GR theories. 

From the first LIGO detections, we know that deviations from GR are presently not detectable.  It is reasonable to assume that this is because any such deviations are less than about a 4\%
effect. While it is possible that the signal-to-noise ratio from the merger itself is currently too small to rule out larger deviations at the horizon, we will not consider this possibility here. Accordingly, rather than simulating black holes in a full quantum theory of gravity, we can consider \textit{effective field theories}. These modify the classical Einstein-Hilbert action of GR through the inclusion of classical terms encompassing quantum gravity effects. One such theory is dynamical Chern-Simons (dCS) gravity, which adds a scalar field coupled to spacetime curvature to the Einstein-Hilbert action, and has origins in string theory, loop quantum gravity, and inflation~\cite{Alexander:2009tp, Green:1984sg, Taveras:2008yf, Mercuri:2009zt, Weinberg:2008hq}. 

The well-posedness of the initial value problem in full, non-linear dCS gravity is unknown~\cite{Delsate:2014hba}. However, we can work in an \textit{order-reduction scheme}, in which we perturb the dCS scalar field and metric about a GR background. At each order, the equations of motion are well-posed (cf.~\cite{Okounkova:2017yby}). In this study, we investigate the behavior of the leading-order dCS metric perturbation, sourced by the leading-order dCS scalar field coupled to the spacetime curvature of a GR background. 

The stability of rotating black holes in dCS gravity is unknown~\cite{Molina:2010fb, Garfinkle:2010zx,Berti:2015itd}. In this study, we numerically test the leading-order stability of rotating dCS black holes by evolving the leading-order dCS metric perturbation on a rotating black hole GR background. Since the background (and the leading-order dCS scalar field) are stationary, the dCS metric perturbation should remain stationary if rotating dCS black holes are stable. 

This question of stability is of broader importance to our goal of simulating the leading-order dCS metric perturbation of a binary black hole spacetime, in order to produce beyond-GR gravitational waveforms. If rotating black holes in dCS are not stable to leading order, and the metric perturbation grows in time, then we know that we would not be able to simulate black hole binaries in this theory. Specifically, the metric perturbations around each black hole would grow in time during inspiral, and similarly for the final black hole after merger, thus spoiling the evolution.

\subsection{Roadmap and conventions} 

This paper is organized as follows. In Sec.~\ref{sec:dCSIntroduction}, we present the equations of motion of dCS that we aim to evolve in this study. In Sec.~\ref{sec:Evolution}, we derive and present a formalism for stably evolving linear metric perturbations on an arbitrary background, so that we may evolve the leading-order dCS metric perturbation. In Sec.~\ref{sec:dCSEvolution}, we apply this formalism to evolve the leading-order dCS metric perturbation on a rotating black hole background. We discuss our findings in Sec.~\ref{sec:Conclusion}.

We set $G = c = 1$ throughout. Quantities are given in terms of  units of $M$, the ADM mass of the background. Latin letters in the beginning of the alphabet $\{a, b,  c, d \ldots \}$ denote 4-dimensional spacetime indices, while Latin letters in the middle of the alphabet $\{i, j ,k, l, \ldots \}$ denote 3-dimensional spatial indices. $g_{ab}$ refers to the spacetime metric, while $\gamma_{ij}$ refers to the spatial metric from a 3+1 decomposition with corresponding timelike unit normal one-form $n_a$ (cf.~\cite{baumgarteShapiroBook} for a review of the 3+1 ADM formalism). 


\section{Dynamical Chern-Simons gravity}
\label{sec:dCSIntroduction}

Dynamical Chern-Simons gravity modifies the Einstein-Hilbert action of GR through the inclusion of a scalar field $\vartheta$, coupled to spacetime curvature as

\begin{align}
\label{eq:dCSAction}
S \equiv \int d^4 x \sqrt{-g} \left( \frac{\mpl^2}{2} R - \frac{1}{2} (\pd \vartheta)^2 - \frac{\mpl}{8} \ell^2 \vartheta \dual RR \right) \,.
\end{align}

The first term in the action is the familiar Einstein-Hilbert action of general relativity, with the Planck mass denoted by $\mpl$. The second term in the action is a kinetic term for the scalar field. The third term, meanwhile, couples $\vartheta$ to spacetime curvature via the Pontryagin density, 
\begin{align}
\dual RR \equiv \dual R^{abcd} R_{abcd}\,,
\end{align}
where $\dual R^{abcd} = \frac{1}{2} \epsilon^{abef} R_{ef}{}^{cd}$ is the dual of the Riemann tensor, and $\epsilon^{abcd} \equiv -[abcd]/\sqrt{-g}$ is the fully antisymmetric Levi-Civita tensor. This coupling is governed by a coupling constant $\ell$, which has dimensions of length. $\ell$ physically represents the length scale below which quantum gravity effects become important. One may also include stress-energy terms in this action for additional fields (such as matter terms in a neutron-star spacetime, for example), though we do not write them here.  

Varying the dCS action with respect to $\vartheta$ gives a sourced wave equation for the scalar field, 
\begin{align}
\square \vartheta = \frac{\mpl}{8} \ell^2 \dual RR\,,
\end{align}
where $\square = \nabla_a \nabla^a$ is the d'Alembertian operator. Varying the action with respect to the metric $g_{ab}$ gives
\begin{align}
\label{eq:MetricEOM}
    \mpl^2 G_{ab} + \mpl \ell^2 C_{ab} = T_{ab}^\vartheta \,,
\end{align}
where 
\begin{align}
\label{eq:CDefinition}
C_{ab} \equiv \epsilon_{cde(a} \nabla^d R_{b)}{}^c \nabla^e \vartheta + \dual R^c{}_{(ab)}{}^d \nabla_c \nabla_d \vartheta \,,
\end{align}
and $T_{ab}^\vartheta$ is the stress energy tensor for a canonical, massless Klein-Gordon field 
\begin{align}
T_{ab}^\vartheta &= \nabla_a \vartheta \nabla_b \vartheta - \frac{1}{2} g_{ab} \nabla_c \vartheta \nabla^c  \vartheta\,.
\end{align}
\textit{It is the inclusion of $C_{ab}$} in Eq.~\eqref{eq:MetricEOM} that modifies the equation of motion for the metric from that of a metric in GR sourced by a scalar field. 

$C_{ab}$, as given in Eq.~\eqref{eq:CDefinition}, contains third derivatives of the metric, thus modifying the principal part of the equation of motion for $\gamma_{ab}$ from that of GR. Because of the presence of these third-derivative terms, it is unknown whether dCS has a well-posed initial value formulation~\cite{Delsate:2014hba}.

However, one can expand the scalar field and metric about a GR background as
\begin{align}
g_{ab} &= g_{ab}^0 + \sum_{k = 1}^\infty \eps^k h_{ab}^{(k)} \,, \\
\vartheta &= \sum_{k = 0}^\infty \eps^k \vartheta^{(k)} \,,
\end{align}
where $\eps$ is an order-counting parameter. At each order in $\eps$ one recovers an equation of motion with the same principal part as GR. This is known as an \textit{order-reduction scheme}, and has been previously implemented in~\cite{Okounkova:2017yby} and~\cite{MashaIDPaper}.

In this scheme, $\eps^0$ simply gives the Einstein field equations of general relativity for $g_{ab}^{(0)}$, with no source term for $\vartheta^{(0)}$, which we can thus set to zero. At first order, we obtain a wave equation for the leading-order scalar field,
\begin{align}
    \square^{(0)} \vartheta^{(1)} = \dual RR^{(0)}\,,
\end{align}
where $\square^{(0)}$ is the d'Alembertian operator of the background, and $RR^{(0)}$ is the Pontryagin density of the background. At this order, the metric perturbation $h_{ab}^{(1)}$ is unsourced, and thus we set it to zero. At order $\eps^2$, the metric perturbation $h_{ab}^{(2)}$ is sourced by the leading-order scalar field $\vartheta^{(1)}$ coupled to spacetime curvature as
\begin{align}
\label{eq:SecondOrder}
    \mpl^2 G_{ab}^{(0)} [h_{ab}^{(2)}] = -\mpl \ell^2 C_{ab}^{(1)} \vartheta^{(1)} + \frac{1}{8} T_{ab}^{(\vartheta(1))} \,,
\end{align}
where $G_{ab}^{(0)}$ is the Einstein field equation operator of the background, and 
\begin{align}
    T_{ab}^{(\vartheta(1))} &\equiv \nabla_a  {}^{(0)} \vartheta^{(1)} \nabla_b {}^{(0)} \vartheta^{(1)} - \frac{1}{2} g_{ab}^{(0)} \nabla_c {}^{(0)} \vartheta^{(1)} \nabla^c {}^{(0)} \vartheta^{(1)} \,,
\end{align}
where $\nabla_a {}^{(0)}$ denotes the covariant derivative associated with $g_{ab}^{(0)}$. Meanwhile, 
\begin{align}
C_{ab}^{(1)}  &\equiv \epsilon_{cde(a} \nabla^d{}^{(0)} R_{b)}{}^c{}^{(0)} \nabla^e {}^{(0)} \vartheta^{(1)} \\
\nn & \quad + \dual R^c{}_{(ab)}{}^d  {}^{(0)} \nabla_c {}^{(0)} \nabla_d {}^{(0)} \vartheta^{(1)} \,.
\end{align}
Note that though $C_{ab}^{(1)}$ contains third derivatives of the background metric $g_{ab}^{(0)}$, it does not contain derivatives of $h_{ab}^{(2)}$, and hence does not contribute to the principal part of Eq.~\eqref{eq:SecondOrder}. We can thus write the RHS of Eq.~\eqref{eq:SecondOrder} in terms of an effective stress energy tensor, 
\begin{align}
    T^{\mathrm{eff}}_{ab}{}^{(1)} &\equiv -\mpl \ell^2 C_{ab}^{(1)} \vartheta^{(1)} + \frac{1}{8} T_{ab}^{(\vartheta(1))}\,.
\end{align}
Let us write Eq.~\eqref{eq:SecondOrder} in a more illuminating way, as
\begin{align}
\label{eq:secorder}
 &\mpl^2 G_{ab}^{(0)} [h_{ab}^{(2)}] = \frac{1}{8} T_{ab}^{(\vartheta(1))}  \\
 \nn & \quad -\mpl \ell^2 \Big( 
 \epsilon_{cde(a} \nabla^d{}^{(0)} R_{b)}{}^c{}^{(0)} \nabla^e {}^{(0)} \vartheta^{(1)} \\
 \nn & \quad + \dual R^c{}_{(ab)}{}^d  {}^{(0)} \nabla_c {}^{(0)} \nabla_d {}^{(0)} \vartheta^{(1)} \Big) \vartheta^{(1)} \,.
\end{align}
As mentioned previously, it is the inclusion of the second term on the right-hand side of Eq.~\eqref{eq:secorder} that differentiates the equation of motion for the leading-order metric perturbation in dynamical Chern-Simons theory from that of a simple metric perturbation sourced by a scalar field in general relativity. 

Our goal, thus, is to evolve the leading-order metric perturbation $h_{ab}^{(2)}$, sourced by $T_{ab}^\mathrm{eff}{}^{(1)}$. Because this is the leading-order metric perturbation, we only need to work in linear theory. We will thus develop a numerical scheme for stably evolving first-order metric perturbations on an arbitrary GR background with arbitrary source. 

From here on, we simplify the notation, writing 
\begin{align}
h_{ab}^{(2)} \equiv \frac{\ell^4}{8} \Delta g_{ab}\,, \; \; \; \vartheta^{(1)} \equiv \frac{\mpl}{8} \ell^2 \Psi\,,
\end{align}
and thus
\begin{align}
\label{eq:CodeTeff}
T^{\mathrm{eff}}_{ab}(\Psi) &\equiv -C_{ab} (\Psi) + \frac{1}{8}T_{ab} (\Psi)\,, \\
\label{eq:CodeC}
C_{ab}(\Psi) &\equiv \epsilon_{cde(a} \nabla^d R_{b)}{}^c \nabla^e \Psi + \dual R^c{}_{(ab)}{}^d \nabla_c \nabla_d \Psi \,, \\
\label{eq:CodeTPsi}
T_{ab}(\Psi) &= \nabla_a \Psi \nabla_b \Psi - \frac{1}{2} g_{ab} \nabla_c \Psi \nabla^c  \Psi \,,
\end{align}
with the overall evolution equation
\begin{align}
\label{eq:CodeEOM}
G_{ab}^{(1)}[\Delta g_{ab}] = T^{\mathrm{eff}}_{ab}(\Psi)\,. 
\end{align}

\section{Evolving metric perturbations}
\label{sec:Evolution}
Our goal now is to outline a formalism to evolve the leading-order metric perturbation in dCS, following Eq.~\eqref{eq:CodeEOM}. In this section, we derive a more general formalism for evolving leading-order metric perturbations on an arbitrary GR background with arbitrary source, which we will apply to rotating black holes in dCS in Sec.~\ref{sec:dCSEvolution}.

\subsection{Generalized harmonic formalism} 
\label{sec:gh}

The formalism that we will use to evolve metric perturbations is based on the generalized harmonic formalism~\cite{Lindblom2006}.  This formulation is a generalization of the well-known harmonic formulation of Einstein's equations, and has seen great success in evolving binary black hole mergers~\cite{friedrich1985, Pretorius:2004jg, Pretorius:2005gq, Lindblom2006}. This well-posed formalism involves  expressing the gauge freedom in terms of a (nearly) freely specifiable gauge source function
\begin{align}
\label{eq:GH}
H_a = g_{ab}\nabla_c \nabla^c x^b = -\Gamma_a\,,
\end{align}
where $\Gamma_a = g^{bc} \Gamma_{abc}$ for the Christoffel symbol derived from $g_{ab}$, and $\nabla_c$ is the corresponding spacetime covariant derivative. Here, $H_a$ is known as the gauge source function, and is a fixed function of coordinates $x^a$ and $g_{ab}$ (but not derivatives of $g_{ab}$). In particular, setting $H_a = 0$ corresponds to a harmonic gauge. This framework has seen success in numerical relativity, including the simulation of black hole binaries~\cite{Pretorius:2004jg, Pretorius:2005gq, Scheel:2008rj}.

In this study, we will consider the first-order formulation of the generalized harmonic formalism given in~\cite{Lindblom2006}. This involves evolving the spacetime metric $g_{ab}$, along with variables $\Pi_{ab}$  and $\Phi_{iab}$ corresponding to its time and spatial derivatives defined as 
\begin{align}
\Phi_{iab} &\equiv \pd_i g_{ab}\,, \\
\Pi_{ab} &\equiv -n^c \pd_c g_{ab}\,,
\end{align}
where $n^c$ is the timelike unit normal vector to slices of constant time $t$. 

For simplicity, we will combine these into a single 4-dimensional variable $\kappa_{abc}$, defined as 
\begin{align}
\label{eq:kappa0}
\kappa_{0ab} &\equiv \Pi_{ab} = -n^c \pd_c g_{ab} \,, \\
\label{eq:kappai}
\kappa_{iab} &\equiv \Phi_{iab} = \pd_i g_{ab} \,.
\end{align}
Note that $\kappa_{abc}$ does not obey the tensor transformation law. 

In addition to being first order, the formalism given in~\cite{Lindblom2006} is also \textit{constraint-damping}. It includes terms proportional to $\pd_i g_{ab} - \kappa_{iab}$, for example; these terms are chosen so that small violations of constraints are driven toward zero. Here, $\pd_i g_{ab}$ is the derivative of $g_{ab}$ taken numerically, while $\kappa_{iab}$ is the first-order variable corresponding to the spatial derivative of the metric. Terms are added to the evolution equations with (spatially-dependent) multiplicative constants $\gamma_0, \gamma_1, \gamma_2$ to ensure symmetric-hyperbolicity and that the relations in Eqs.~\eqref{eq:GH},~\eqref{eq:kappa0} and~\eqref{eq:kappai} are obeyed. 

The first-order, symmetric-hyperbolic, constraint-damping evolution equations for the metric are given by
\begin{align}
\label{eq:dtpsi}
\pd_t g_{ab} &= (1 + \gamma_1)\beta^k \pd_k g_{ab} - \alpha \kappa_{0ab} - \gamma_1 \beta^i \kappa_{iab}\,, \\
\label{eq:dtkappai}
\pd_t \kappa_{iab} &= \beta^k \pd_k \kappa_{iab} - \alpha \pd_i \kappa_{0ab} + \alpha \gamma_2 \pd_i g_{ab} - \alpha \gamma_2 \kappa_{iab} \\
\nonumber &\quad+ \frac{1}{2} \alpha n^c n^d \kappa_{icd} \kappa_{0ab} + \alpha \gamma^{jk} n^c \kappa_{ijc} \kappa_{kab}  \,, 
\end{align}
and
\begin{align}
\label{eq:dtkappa0}
\pd_t \kappa_{0ab} &= \beta^k \pd_k \kappa_{0ab} - \alpha \gamma^{ki} \pd_k \kappa_{iab} + \gamma_1 \gamma_2 \beta^k \pd_k g_{ab} \\
\nonumber &+ 2 \alpha g^{cd} (\gamma^{ij} \kappa_{ica} \kappa_{jdb} - \kappa_{0ca} \kappa_{0db} \\
\nn &- g^{ef} \Gamma_{ace} \Gamma_{bdf} ) \\
\nonumber &- 2 \alpha \nabla_{(a} H_{b)} - \frac{1}{2} \alpha n^c n^d \kappa_{0cd} \kappa_{0ab} \\
\nonumber & - \alpha n^c \kappa_{0ci} \gamma^{ij} \kappa_{jab} \\
\nonumber &+ \alpha \gamma_0 [2 \delta^c{}_{(a}n_{b)} - g_{ab}n^c](H_c + \Gamma_c) \\
\nonumber &- \gamma_1 \gamma_2 \beta^i \kappa_{iab} - 2 \alpha S_{ab}\,. 
\end{align}
In the last equation, $S_{ab}$ is a source term related to trace-reverse of the stress-energy tensor $T_{ab}$ as 
\begin{align}
\label{eq:Sab}
S_{ab} = 8 \pi (T_{ab} - \frac{1}{2} T g_{ab}) \,,
\end{align}
where $T = g^{ab} T_{ab}$. In the above, $\nabla_a H_b$ is defined as $\partial_a H_b - \Gamma^d{}_{ab} H_d$, as if $H_a$ were a one-form (which it is not).
\subsection{Linearized generalized harmonic formalism} 
\label{sec:variables} 

Our goal in this study is to evolve first-order metric perturbations on a GR background. Given a background $\{g_{ab}, \kappa_{abc}\}$, we perturb it to first order as
\begin{align}
g_{ab} &\to g_{ab} + \Delta g_{ab} \,, \\
\kappa_{abc} &\to \kappa_{abc} + \Delta \kappa_{abc} \,.
\end{align}
From here on, $\Delta A$ will always refer to the linear perturbation to a variable $A$. 

The evolution equations for $\Delta g_{ab}$ and $\Delta \kappa_{abc}$ can be derived by linearizing Eqs.~\eqref{eq:dtpsi},~\eqref{eq:dtkappai} and~\eqref{eq:dtkappa0}, and keeping terms to first order. The resulting equations will be a first-order formulation. The 
symmetric hyperbolicity of these equations is guaranteed because the perturbation equations will have the same principal part as the background system. The linearized system is also constraint damping, as the associated \textit{constraint evolution system} has the same linear part as in the constraint-damping unperturbed system (cf. Eqs.~17 -- 21 in~\cite{Lindblom2006}). More importantly, the equations for $\Delta g_{ab}$ and $\Delta \kappa_{abc}$ will have the same principal part as the equations for $g_{ab}$ and $\kappa_{abc}$, as we shall see. 

Linearizing Eqs.~\eqref{eq:dtpsi},~\eqref{eq:dtkappai} and~\eqref{eq:dtkappa0} involves computing terms like $\Delta \alpha$, $\Delta \beta^i$, the first-order perturbations to the lapse and shift. In the following section, we thus derive expressions for these terms in terms of the fundamental variables $\Delta g_{ab}$ and $\Delta \kappa_{abc}$. 

\subsection{Linearized variables}

To compute $\Delta g^{ab}$, we can use the 
identity $g^{ab} g_{bc} = \delta^a_c$ to give 
\begin{align}
\label{eq:hpsiupper}
\Delta g^{ad} = -  g^{cd} g^{ab} \Delta g_{bc}\,.
\end{align}

For the perturbation to the lapse, $\Delta \alpha$, the shift, $\Delta \beta^i$, the lower-indexed shift, $\Delta \beta_i$, and the spatial metric $\Delta \gamma_{ij}$ and $\Delta \gamma^{ij}$, we recall 
that the spacetime metric is decomposed in the 3+1 ADM formalism as
\begin{align}
g_{ab} &= \left( \begin{smallmatrix} -\alpha^2 + \beta_l \beta^l & \beta_i \\ \beta_j & \gamma_{ij} \end{smallmatrix} \right) \\
g^{ab} &= \left( \begin{smallmatrix} -\alpha^{-2} & \alpha^{-2} \beta^i \\ \alpha^{-2} \beta^j & \gamma^{ij} - \alpha^{-2} \beta^i \beta^j \end{smallmatrix} \right).
\end{align} 
Recall that spatial quantities are raised and lowered with $\gamma_{ij}$, the spatial metric. When we perturb all 10 independent components of $g_{ab}$, we can find what all of the linearized quantities are in terms 
of $g_{ab}$ and $\Delta g_{ab}$. We begin with perturbing $g_{0i}$ to find $\Delta \beta_i$:
\begin{align}
\label{eq:hbetalower}
\Delta \beta_i = \Delta g_{0i}\,.
\end{align}
Similarly, we can perturb $g_{ij}$
to obtain:
\begin{align}
\label{eq:hglower}
\Delta \gamma_{ij} &= \Delta g_{ij}\,.
\end{align}
We can now use $g^{00}$ to obtain
\begin{align}
\label{eq:halpha}
 \Delta \alpha = \frac{1}{2}\alpha^3 \Delta g^{00}\,.
\end{align}
Next, using $\gamma^{ij} \gamma_{jk} = \delta^i_k$, we find
\begin{align}
\Delta \gamma^{im} &= - \gamma^{mk} \gamma^{ij} \Delta \gamma_{jk} \,.
\end{align} 
From this, we can compute $\Delta \beta^i$ as 
\begin{align}
\Delta \beta^i = \Delta \gamma^{ij} \beta_j + \gamma^{ij} \Delta \beta_j \,.
\end{align}

Finally, we need to compute $\Delta n^a$ and $\Delta n_a$, the perturbed time-like unit normal vector and one-form. We can use the expressions for $n^a$ and $n_a$ in terms of the lapse and shift to obtain the perturbed quantities (cf.~\cite{baumgarteShapiroBook}). We compute
\begin{align}
\label{eq:hnlower}
\Delta n_a& = (-\Delta \alpha, 0, 0, 0)\,.
\end{align}
and
\begin{align}
\label{eq:hnupper}
\Delta n^a &= (-\alpha^{-2}\Delta \alpha, \alpha^{-2}\Delta \alpha\beta^i - \alpha^{-1}\Delta \beta^i)\,.
\end{align}

In order to check constraint satisfaction (as will be discussed in Sec.~\ref{sec:Constraints}), we will also need to obtain the perturbation to $\gamma_{a}^b$. We obtain (cf. Eq.~2.30 in~\cite{baumgarteShapiroBook}),
\begin{align}
\Delta \gamma^{a}{}_{b} &= \Delta n^a n_b + n^a \Delta n_b\,.
\end{align}

Thus, we have obtained all of the necessary perturbed quantities to perturb the generalized harmonic expressions as well as the constraint expressions that we can obtain from $\Delta g_{ab}$. In the next section, we describe the quantities that we can obtain from $\Delta \kappa_{abc}$.

Referring back to Eq.~\eqref{eq:dtkappa0}, we also need to find expressions for $\Delta \Gamma_{abc}$, the first-order perturbation to the connection compatible with $g_{ab}$, 
as well as the first-order perturbation to its trace, $\Delta \Gamma_a$. First, let's compute the perturbation to $\Delta \Gamma_{abc}$. By definition,
\begin{align}
\Gamma_{abc} = \frac{1}{2}(\pd_b g_{ac} + \pd_c g_{ab} - \pd_a g_{bc})\,.
\end{align}
However, in order to preserve hyperbolicity in the evolution equations, all instances of $\pd_a g_{bc}$ appearing in $\Gamma_{abc}$ are replaced with $\kappa_{abc}$ according to Eqs.~\eqref{eq:kappa0} and~\eqref{eq:kappai}~\cite{Lindblom2006}, thus giving
\begin{align}
\label{eq:christoffel}
    \Gamma_{abc} = \frac{1}{2}\Big(&(1 - \delta_b^0) \kappa_{bac} + \delta_b^0(-\alpha \kappa_{0ac} + \beta^i \kappa_{iac}) \\
    \nn \quad +\,& (1 - \delta_c^0) \kappa_{cab} + \delta_c^0(-\alpha \kappa_{0ab} + \beta^i \kappa_{iab}) \\
    \nn \quad -\,& (1 - \delta_a^0) \kappa_{abc} - \delta_a^0(-\alpha \kappa_{0bc} + \beta^i \kappa_{ibc}) \Big) 
\end{align}
where the Kronecker  delta symbol $\delta^a_b$ picks out the spatial indices $\{1,2,3\}$ versus time indices $\{0\}$.

We can perturb Eq.~\eqref{eq:christoffel} to give
\begin{align}
\label{eq:hchristoffel}
    &\Delta\Gamma_{abc} = \frac{1}{2}\Big((1 - \delta_b^0) \Delta \kappa_{bac} \\
    \nn &\quad + \delta_b^0 (-\Delta \alpha \kappa_{0ac} -\alpha \Delta \kappa_{0ac} + \Delta \beta^i \kappa_{iac} + \beta^i \Delta \kappa_{iac}) \\
    \nn &\quad + (1 - \delta_c^0) \Delta \kappa_{cab} \\
    \nn &\quad + \delta_c^0(-\Delta \alpha \kappa_{0ab} -\alpha \Delta \kappa_{0ab} + \Delta \beta^i \kappa_{iab} + \beta^i \Delta \kappa_{iab}) \\
    \nn &\quad -(1 - \delta_a^0) \Delta \kappa_{abc} \\
    \nn &\quad - \delta_a^0(-\Delta \alpha \kappa_{0bc} -\alpha \Delta \kappa_{0bc} + \Delta \beta^i \kappa_{ibc} + \beta^i \Delta \kappa_{ibc}) \Big) 
\end{align}


Now, for $\Gamma^a{}_{bc} \equiv g^{ad} \Gamma_{dbc}$, we compute the corresponding perturbations (for future use) via
\begin{align}
\label{eq:hChristoffel2ndKind}
\Delta \Gamma^a{}_{bc} = \Delta g^{ad} \Gamma_{dbc} + g^{ad} \Delta \Gamma_{dbc}\,.
\end{align}
For the trace of $\Gamma_a \equiv g^{bc} \Gamma_{abc}$, we compute
\begin{align}
\label{eq:hGammaA}
\Delta \Gamma_a &= \Delta g^{bc} \Gamma_{abc} + g^{bc} \Delta \Gamma_{abc}\,,
\end{align}
where $\Delta \Gamma_{abc}$ is as above, and $\Delta g^{bc}$ is given in Eq.~\eqref{eq:hpsiupper}.

The generalized harmonic gauge source term, $H_a$, will also have a perturbation, $\Delta H_a$. However, $\Delta H_a$, like $H_a$, is freely specifiable, with the caveat that it can only depend on $g_{ab}$ and $\Delta g_{ab}$ but no derivatives of $g_{ab}$ or $\Delta g_{ab}$. Throughout this study we will choose a \textit{freezing} gauge condition: we set $\Delta H_a$ from the initial data $\Delta H_a = \Delta \Gamma_a (t = 0)$, and keep  it at this constant value throughout the evolution. 

Eq.~\eqref{eq:dtkappa0} has a $\nabla_a H_b$ term. Perturbing this quantity, we obtain
\begin{align}
\label{eq:hdelH}
\Delta ( \nabla_a H_b) &= \pd_a \Delta H_b - \Delta g^{cd}\Gamma_{dab} H_c \\
\nn &\quad - g^{cd} (\Delta \Gamma_{dab} H_c + \Gamma_{dab} \Delta H_c) \,.
\end{align} 

\subsubsection{Perturbed initial data}
Suppose we are given initial data in the form $\{\Delta g_{ab}, \pd_t \Delta g_{ab}, \pd_i \Delta g_{ab}\}$. Perturbing Eqs.~\eqref{eq:kappa0} and~\eqref{eq:kappai}, we can relate $\Delta \kappa_{abc}$ to derivatives of $\Delta g_{ab}$: 

\begin{align}
\label{eq:kappa0Init}
\Delta \kappa_{0ab} &= -\Delta n^c \pd_c g_{ab} -n^c \pd_c \Delta g_{ab}\,, \\
\label{eq:kappaiInit}
\Delta \kappa_{iab} &= \pd_i \Delta g_{ab} \,,
\end{align}
where $\Delta n^c$ is computed from $\Delta g_{ab}$ using Eq.~\eqref{eq:hnupper}. 

\subsubsection{Source terms}
\label{sec:SourceTerms}

In order to source the metric perturbation, we require a perturbation to the stress energy tensor, $\Delta T_{ab}$. This will appear in the 
perturbed evolution equations through $\Delta S_{ab}$, the perturbation to $S_{ab}$ defined in Eq.~\eqref{eq:Sab}, as 
\begin{align}
\label{eq:hSab}
\Delta S_{ab} &= 8 \pi(\Delta T_{ab} - \frac{1}{2}(\Delta T g_{ab} + T \Delta g_{ab})) \,, \\
\Delta T &= \Delta g^{ab} T_{ab} + g^{ab} \Delta T_{ab}\,.
\end{align}
For a vacuum background, we obtain the simpler form
\begin{align}
\Delta S_{ab} = 8 \pi(\Delta T_{ab} - \frac{1}{2} g_{ab} g^{cd} \Delta T_{cd} ) \,.
\end{align}

\subsection{Perturbed evolution equations}
\label{sec:EvolutionEquations}

We have now derived the first-order perturbations to all of the variables in Eqs.~\eqref{eq:dtpsi},~\eqref{eq:dtkappai}, and~\eqref{eq:dtkappa0}. We next perturb these equations to linear order, in order to obtain evolution equations for $\Delta g_{ab}$ and $\Delta \kappa_{abc}$.

We begin by perturbing Eq.~\eqref{eq:dtpsi} to obtain
\begin{align}
\label{eq:dthpsi}
\pd_t \Delta g_{ab} &= (1 + \gamma_1)(\Delta \beta^k \pd_k g_{ab} + \beta^k \pd_k \Delta g_{ab}) \\
\nn &\quad - \Delta \alpha  \kappa_{0ab}  - \alpha \Delta \kappa_{0ab} \\
\nn &\quad - \gamma_1 \Delta \beta^i \kappa_{iab} - \gamma_1 \beta^i \Delta \kappa_{iab}\,.
\end{align}
Next, we perturb Eq.~\eqref{eq:dtkappai} to give
\begin{align}
\label{eq:dthkappaiab}
\pd_t \Delta &\kappa_{iab} = \Delta \beta^k \pd_k \kappa_{iab} + \beta^k \pd_k \Delta \kappa_{iab} \\
\nn &\quad  - \Delta \alpha \pd_i \kappa_{0ab} - \alpha \pd_i \Delta \kappa_{0ab} \\
\nn&\quad + \Delta \alpha \gamma_2 \pd_i g_{ab}  + \alpha \gamma_2 \pd_i \Delta g_{ab} \\
\nn&\quad + \frac{1}{2} \Delta \alpha n^c n^d \kappa_{icd} \kappa_{0ab} + \frac{1}{2} \alpha \Delta n^c n^d \kappa_{icd} \kappa_{0ab} \\
\nn&\quad + \frac{1}{2} \alpha n^c \Delta n^d \kappa_{icd} \kappa_{0ab} + \frac{1}{2} \alpha n^c n^d \Delta \kappa_{icd} \kappa_{0ab} \\
\nn&\quad + \frac{1}{2} \alpha n^c n^d \kappa_{icd} \Delta \kappa_{0ab} \\
\nn&\quad + \Delta \alpha \gamma^{jk} n^c \kappa_{ijc} \kappa_{kab} + \alpha \Delta \gamma^{jk} n^c \kappa_{ijc} \kappa_{kab} \\
\nn&\quad + \alpha \gamma^{jk} \Delta n^c \kappa_{ijc} \kappa_{kab} + \alpha \gamma^{jk} n^c \Delta \kappa_{ijc} \kappa_{kab} \\
\nn&\quad + \alpha \gamma^{jk} n^c \kappa_{ijc} \Delta \kappa_{kab} \\
\nn&\quad - \Delta \alpha \gamma_2 \kappa_{iab} - \alpha \gamma_2 \Delta \kappa_{iab}\,.
\end{align}
Finally, we perturb Eq.~\eqref{eq:dtkappa0} to obtain
\begin{align}
\label{eq:dthkappa0ab}
&\pd_t \Delta \kappa_{0ab} =  \Delta \beta^k \pd_k \kappa_{0ab} +  \beta^k \pd_k \Delta \kappa_{0ab} \\
\nn &\quad - \Delta \alpha \gamma^{ki} \pd_k \kappa_{iab}  - \alpha \Delta \gamma^{ki} \pd_k \kappa_{iab} \\
\nn &\quad  - \alpha \gamma^{ki} \pd_k \Delta \kappa_{iab} \\
\nn &\quad + \gamma_1 \gamma_2 \Delta \beta^k \pd_k g_{ab} + \gamma_1 \gamma_2 \beta^k \pd_k \Delta g_{ab} \\
\nn &\quad + 2 \Delta \alpha g^{cd} (\gamma^{ij} \kappa_{ica} \kappa_{jdb} - \kappa_{0ca} \kappa_{0db} - g^{ef} \Gamma_{ace} \Gamma_{bdf} ) \\
\nn &\quad + 2 \alpha \Delta g^{cd} (\gamma^{ij} \kappa_{ica} \kappa_{jdb} - \kappa_{0ca} \kappa_{0db} - g^{ef} \Gamma_{ace} \Gamma_{bdf} ) \\
\nn &\quad + 2 \alpha g^{cd} (\Delta \gamma^{ij} \kappa_{ica} \kappa_{jdb} - \Delta \kappa_{0ca} \kappa_{0db} - \Delta g^{ef} \Gamma_{ace} \Gamma_{bdf} ) \\
\nn &\quad + 2 \alpha g^{cd} (\gamma^{ij} \Delta \kappa_{ica} \kappa_{jdb} - \kappa_{0ca} \Delta \kappa_{0db} - g^{ef} \Delta \Gamma_{ace} \Gamma_{bdf} ) \\
\nn &\quad + 2 \alpha g^{cd} (\gamma^{ij} \kappa_{ica} \Delta \kappa_{jdb} - g^{ef} \Gamma_{ace} \Delta \Gamma_{bdf} ) \\
\nn &\quad - 2 \Delta \alpha \nabla_{(a} H_{b)} - 2 \alpha \Delta\nabla_{(a} H_{b)} \\
\nn &\quad - \frac{1}{2} \Delta \alpha n^c n^d \kappa_{0cd} \kappa_{0ab}- \frac{1}{2} \alpha \Delta n^c n^d \kappa_{0cd} \kappa_{0ab}\\
\nn &\quad - \frac{1}{2} \alpha n^c \Delta n^d \kappa_{0cd} \kappa_{0ab}- \frac{1}{2} \alpha n^c n^d \Delta \kappa_{0cd} \kappa_{0ab}\\
\nn &\quad - \frac{1}{2} \alpha n^c n^d \kappa_{0cd} \Delta \kappa_{0ab} \\
\nn &\quad - \Delta \alpha n^c \kappa_{0ci} \gamma^{ij} \kappa_{jab} - \alpha \Delta n^c \kappa_{0ci} \gamma^{ij} \kappa_{jab} \\
\nn &\quad - \alpha n^c \Delta \kappa_{0ci} \gamma^{ij} \kappa_{jab} - \alpha n^c \kappa_{0ci} \Delta \gamma^{ij} \kappa_{jab}\\
\nn &\quad - \alpha n^c \kappa_{0ci} \gamma^{ij} \Delta \kappa_{jab} \\
\nn &\quad + \Delta \alpha \gamma_0 [2 \delta^c{}_{(a}n_{b)} - g_{ab}n^c](H_c + \Gamma_c) \\
\nn &\quad + \alpha \gamma_0 [2 \delta^c{}_{(a}\Delta n_{b)} - \Delta g_{ab}n^c](H_c + \Gamma_c) \\
\nn &\quad + \alpha \gamma_0 [- g_{ab}\Delta n^c](H_c + \Gamma_c) \\
\nn &\quad + \alpha \gamma_0 [2 \delta^c{}_{(a}n_{b)} - g_{ab}n^c](\Delta H_c + \Delta \Gamma_c) \\
\nn &\quad - \gamma_1 \gamma_2 \Delta \beta^i \kappa_{iab}  - \gamma_1 \gamma_2 \beta^i \Delta \kappa_{iab} \\
\nn &\quad -2 \Delta \alpha S_{ab} - 2 \alpha \Delta S_{ab} \,.
\end{align}

\subsection{Constraint Equations}
\label{sec:Constraints}

In order to check the numerical performance of the evolution equations given in the previous section, we evaluate a set of four perturbed constraints that $\Delta g_{ab}$ and $\Delta \kappa_{abc}$ must satisfy. These functions are zero analytically, and we will check their convergence to zero with increasing numerical resolution. 

The 1-index constraint (cf.~\cite{Lindblom2006}) is the gauge constraint 
\begin{align}
\label{eq:1con}
C_a &= H_a + \Gamma_a\,,
\end{align}
which measures the numerical accuracy of the generalized harmonic evolution (cf. Eq.~\eqref{eq:GH}). We perturb this to get the constraint
\begin{align}
\label{eq:h1con}
\Delta C_a &\equiv \Delta H_a + \Delta \Gamma_a \,,
\end{align}
where $\Delta H_a$ is the gauge source function for the metric perturbation evolution. 

The 3-index constraint evaluates the difference between the numerical derivative of $g_{ab}$ and $\kappa_{iab}$, the first-order variable encoding the spatial derivative of the metric as
\begin{align}
\label{eq:3con}
C_{iab} &= \pd_i g_{ab} - \kappa_{iab} \,. 
\end{align}
Perturbing this, we obtain
\begin{align}
\label{eq:h3con}
\Delta C_{iab} &= \pd_i \Delta g_{ab} - \Delta \kappa_{iab}\,.
\end{align}

The 4-index constraint concerns the commutation of partial derivatives as
\begin{align}
\label{eq:4con}
C_{ijab} \equiv 2 \pd_{[i}\kappa_{j]ab}\,.
\end{align}
Perturbing this, we obtain
\begin{align}
\label{eq:h4con}
\Delta C_{ijab} \equiv 2 \pd_{[i}\Delta \kappa_{j]ab}\,.
\end{align}

Finally, the 2-index constraint is derived from the Hamiltonian and momentum constraints, as well as the 3-index constraint. The constraint and its perturbation are too lengthy to reproduce here, and so we have written them in Appendix~\ref{sec:2con}.

Thus, when performing an evolution, we evaluate the right-hand sides of Eqs.~\eqref{eq:h1con},~\eqref{eq:h3con},~\eqref{eq:h4con} and~\eqref{eq:h2con}, and check that they converge to zero with increasing numerical resolution. In particular, as we use a spectral code, we expect exponential convergence with resolution~\cite{SpECwebsite}.

In order to show that the constraints themselves are convergent, rather than the absolute values of the metric variables simply getting smaller, we can normalize the constraints by the absolute values of the metric fields they contain. For example, for a constraint of the form $A + B$, we normalize it by dividing by $\sqrt{A^2 + B^2}$. The question arises of whether we should normalize the constraints pointwise, or whether we should compute the norm of each constraint and its normalization factor over the entire domain and then divide the norms. Since we will evolve a localized metric perturbation, there will be regions in the domain with $\Delta g_{ab}$ nearly zero, so we choose to first compute norms and then divide them.

\subsection{Characteristic variables}
\label{sec:CharProjection}

All of the discussion so far has centered on fundamental variables $\Delta g_{ab}$ and $\Delta \kappa_{abc}$. However, in order to implement boundary conditions, it is useful
to instead consider \textit{characteristic fields}. These can be used to measure the characteristic speeds and to construct boundary conditions. 

The characteristic fields are the eigenvectors of the principal part of the evolution equations (cf.~\cite{Lindblom2006} for an example derivation). The characteristic speeds are the corresponding eigenvalues. For the generalized harmonic system, the characteristic variables on a surface with spatial normal vector $\hat{n}^i$ take the form 
\begin{align}
\label{eq:u0}
u^{0}_{ab} &= g_{ab} \,, \\
\label{eq:u1}
u^{1\pm}_{ab} &= \kappa_{0ab} \pm \hat{n}^i \kappa_{iab} - \gamma_2 g_{ab} \,, \\
\label{eq:u2}
u^{2}_{iab} &= (\delta_i{}^k - \hat{n}_i \hat{n}^k)\kappa_{kab}\,.
\end{align} 

The principal parts of the linearized equations (cf. Sec~\ref{sec:EvolutionEquations}) are
\begin{align}
\label{eq:hPP}
&\pd_t \Delta g_{ab} - (1 + \gamma_1) \beta^k \pd_k \Delta g_{ab} \simeq 0 \,, \\
&\pd_t \Delta \kappa_{0ab} - \beta^k \pd_k \Delta \kappa_{0ab} \\
\nn & \quad + \alpha \gamma^{ki} \pd_k \Delta \kappa_{iab} - \gamma_1 \gamma_2 \beta^k \pd_k \Delta g_{ab}  \simeq 0 \,, \\
& \pd_t \Delta \kappa_{iab} - \beta^k \pd_k \Delta \kappa_{iab} \\
\nn & \quad + \alpha \pd_i \Delta \kappa_{0ab} - \gamma_2 \alpha \pd_i \Delta g_{ab} \simeq 0 \,.
\end{align}
These are exactly those of the generalized harmonic system, and hence the characteristic fields and speeds will be the same. Thus, the characteristic fields of the linearized system are simply
\begin{align}
\label{eq:hu0}
\Delta u^{0}_{ab} &= \Delta g_{ab} \,, \\
\label{eq:hu1}
\Delta u^{1\pm}_{ab} &= \Delta \kappa_{0ab} \pm \hat{n}^i \Delta \kappa_{iab} - \gamma_2 \Delta g_{ab} \,, \\
\label{eq:hu2}
\Delta u^{2}_{iab} &= (\delta_i{}^k - \hat{n}_i \hat{n}^k)\Delta \kappa_{kab}\,.
\end{align}

The reverse transformation from characteristic variables to fundamental variables is then
\begin{align}
\Delta g_{ab} &= \Delta u^{0}_{ab} \,, \\
\Delta \kappa_{0ab} &= \frac{1}{2}(\Delta u^{1+}_{ab} + \Delta u^{1-}_{ab}) + \gamma_2 \Delta u^{0}_{ab} \,, \\
\Delta \kappa_{iab} &= \frac{1}{2}\hat{n}_i(\Delta u^{1+}_{ab} - \Delta u^{1-}_{ab}) + \Delta u^{2}_{iab}\,.
\end{align}

As in the generalized harmonic system, the characteristic speed for $\Delta u^0_{ab}$ 
is $-(1 + \gamma_1)n_k\beta^k$, the speed for $\Delta u_{ab}^{1\pm}$ is
$-n_k\beta^k \pm \alpha$, and the speed for $\Delta u_{iab}^{2}$ is $-n_k \beta^k$.

\subsection{Boundary Conditions}
\label{sec:BoundaryConditions}

In the previous section, we derived the characteristic fields for the linearized system. In order to complete the evolution system, we must include boundary conditions for these characteristic fields. All of our numerical evolutions include a finite outer boundary, and we choose to use a freezing boundary condition, which sets
\begin{align}
\label{eq:BoundaryCondition}
P(d \Delta u^{(a)} / dt) = 0 \,,
\end{align}
where $\Delta u^{(a)}$ is a perturbation to a characteristic field and $P$ refers to the characteristic projection onto the surface. Though more sophisticated conditions are available, especially for computing accurate gravitational radiation (cf.~\cite{Kidder:2004rw, Rinne:2007ui, Rinne:2008vn}), we find that the freezing boundary condition is sufficient for our purposes, especially since the characteristics are initially purely outgoing (out of the computational domain). 

When simulating metric perturbations on a spacetime containing one or more black holes, we exclude the region just inside the apparent horizon from the computational domain~\cite{Hemberger:2012jz}. This forms a topologically spherical inner boundary. However, there should be no characteristics entering the computational domain from the horizon, and thus we do not need to specify a condition at the inner boundary. 

\subsection{Code Tests}
\label{sec:CodeTests}

Because of the complexity of Eqs.~\eqref{eq:dtpsi},~\eqref{eq:dtkappai}, and~\eqref{eq:dtkappa0}, we perform a series of code tests. These code tests contain no new physics, but rather check that the evolution equations have been implemented correctly. We present the results of these tests in Appendix~\ref{sec:CodeTestsAppendix}.

\clearpage


\section{Evolving dCS metric perturbations}
\label{sec:dCSEvolution}


We now apply the formalism given in Sec.~\ref{sec:Evolution} to dynamical Chern-Simons gravity. Specifically, we aim to test the stability of rotating black holes in dCS by evolving the leading-order metric perturbation, $\Delta g_{ab}$, governed by Eq.~\eqref{eq:CodeEOM}, on a rotating black hole background. In GR, this background is given by the Kerr metric. Recall from Eqs.~\eqref{eq:CodeTeff},~\eqref{eq:CodeC}, and~\eqref{eq:CodeTPsi}, that it is precisely the inclusion of $C_{ab} (\Psi)$ in the effective stress energy tensor that differentiates dynamical Chern-Simons gravity, where the scalar field is coupled to spacetime curvature via $\dual RR$, from a simple metric perturbation sourced by a scalar field in GR. 

\subsection{Implementation details}

In~\cite{MashaIDPaper}, we derived stationary initial data for $\Delta g_{ab}$ on a Kerr background sourced by the spacetime curvature of the Kerr background coupled to a stationary field $\Psi$ obeying $\square \Psi = \dual RR\,.$~\cite{Stein:2014xba}. Using these data, we construct $\Delta \kappa_{abc}$ following Eqs.~\eqref{eq:kappaiInit} and~\eqref{eq:kappa0Init}. The source term $\Delta S_{ab}$ described in Sec.~\ref{sec:SourceTerms} is computed from $\Psi$ using $T^{\mathrm{eff}}_{ab}(\Psi)$ in Eq.~\eqref{eq:CodeTeff}.

Our computational domain is a set of eleven nested spherical shells, with more shells centered near the horizon and fewer shells further out. The boundary of the innermost shell conforms to the apparent horizon of the background black hole, and the outer boundary is  at $R = 200\,M$.  We repeat simulations at three different numerical resolutions determined by a parameter labeled "low", "medium", or "high"; each shell has 5 radial spectral basis points and 6 angular spectral basis points at the lowest resolution, with one more radial and angular basis point added for each increase in our resolution parameter.

\begin{figure}
  \centering
  \label{fig:EvConstraintDamping}
  \includegraphics[width=\columnwidth]{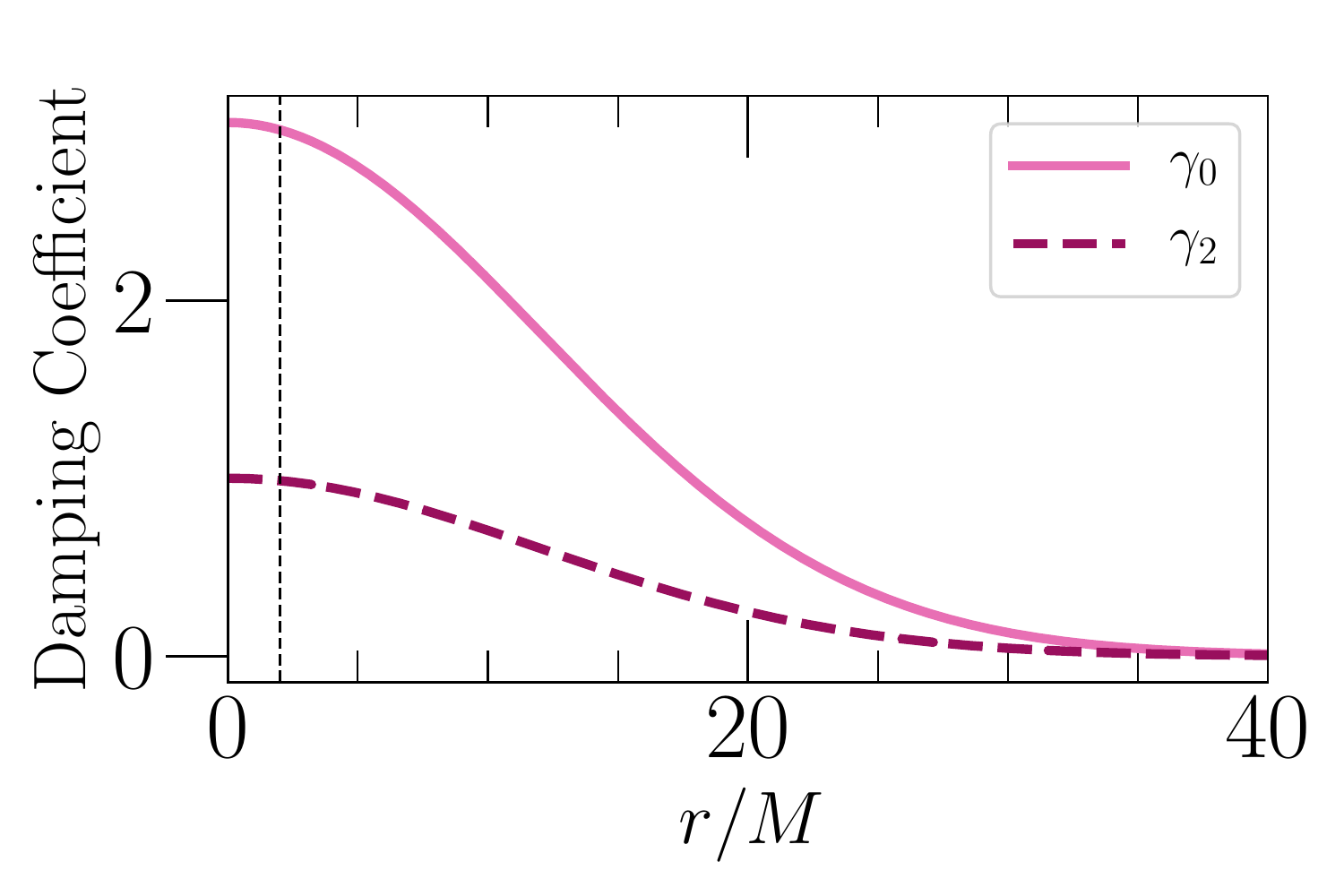}
  \caption{
  Constraint damping functions $\gamma_0$ and $\gamma_2$ used to evolve metric perturbations on a Kerr background. The functions are largest where the metric perturbation source has the highest value, and exponentially decay to $R \to \infty$. While the functions extend to $R = 0$, the computational domain terminates outside the apparent horizon inner boundary (here shown by the black dashed line at $R = 2\,M$ in the case of Schwarzschild). 
  }
\end{figure}

We evolve $\{\Delta g_{ab}, \Delta \kappa_{abc}\}$ using the equations in Sec.~\ref{sec:EvolutionEquations} using a spectral code~\cite{SpECwebsite}. We apply filtering to the spectral scheme in order to minimize the growth of high-frequency modes~\cite{Szilagyi:2009qz}. We choose damping parameters $\gamma_0$ and $\gamma_2$ to be larger close to the horizon, where the metric perturbation is greatest, as shown in Fig.~\ref{fig:EvConstraintDamping}. We choose $\gamma_1 = -1$ as in Ref.~\cite{Lindblom2006}. 

\subsection{Results}

\begin{figure}
  \centering
  \label{fig:Spin0p1EvConstraints}
  \includegraphics[width=\columnwidth]{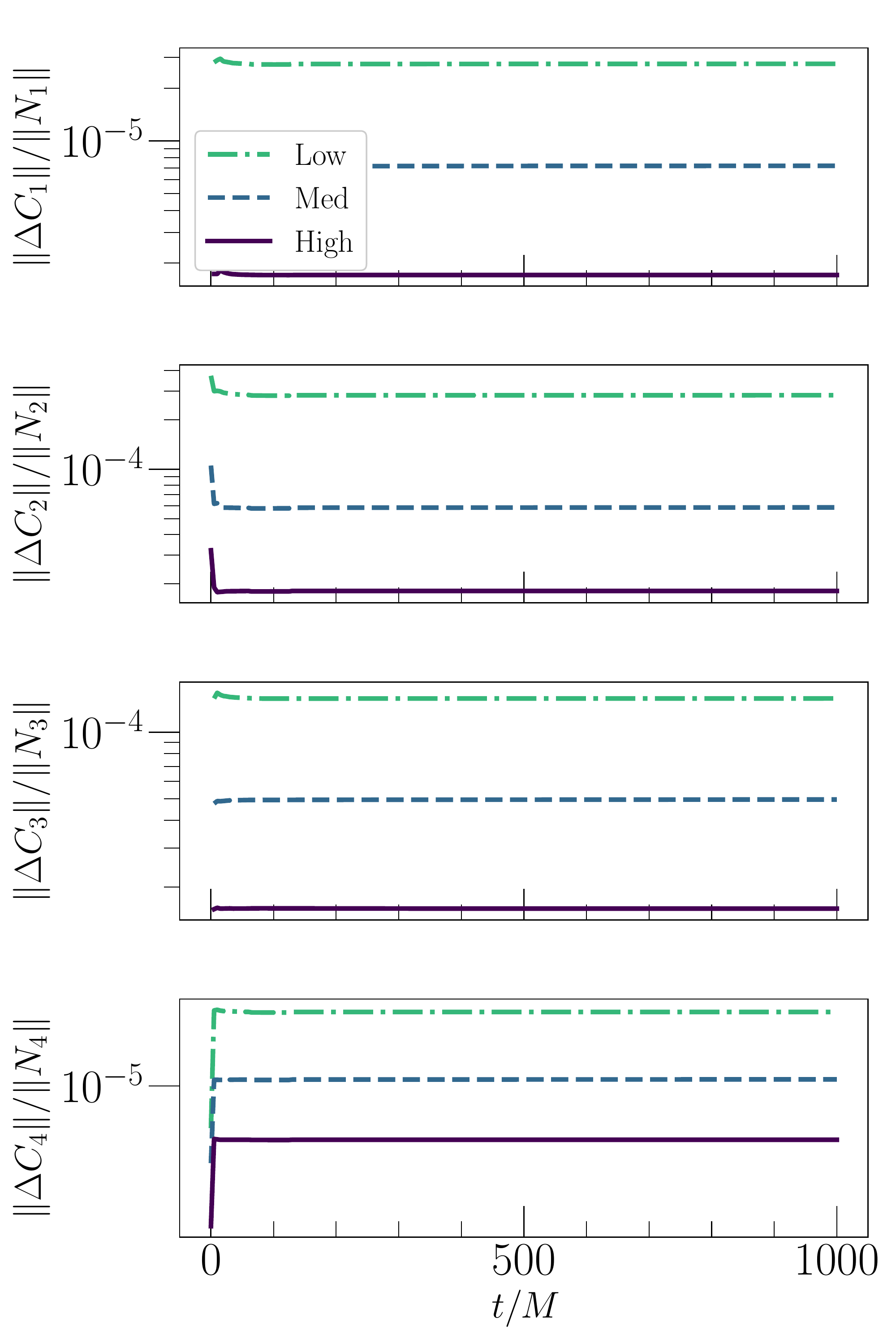}
  \caption{Behavior of the perturbed constraints given in Sec.~\ref{sec:Constraints} for a dCS perturbation on a Kerr background with $\chi = 0.1$. For each constraint $\Delta C_A$, we compute the L2 norm of the constraint over the entire computational domain ($\|\Delta C_1\|$ for the 1-index constraint, for example), and divide by the L2 norm of its normalization factor ($\| N_A \|$) (cf. Sec.~\ref{sec:Constraints}).  We see that the constraints remain constant in time and are exponentially convergent with resolution. 
  }
\end{figure}

In Fig.~\ref{fig:Spin0p1EvConstraints}, we present the perturbed constraint violation for a spin $\chi = 0.1$ background using the expressions derived in Sec.~\ref{sec:Constraints}. We see that the constraints remain roughly constant in time, and are exponentially convergent. We check the constraint convergence for every simulation. Note that as we increase the spin, more spectral coefficients are needed to achieve the same level of constraint violation. 

\begin{figure}
  \centering
  \label{fig:Spin0p1EvVars}
  \includegraphics[width=\columnwidth]{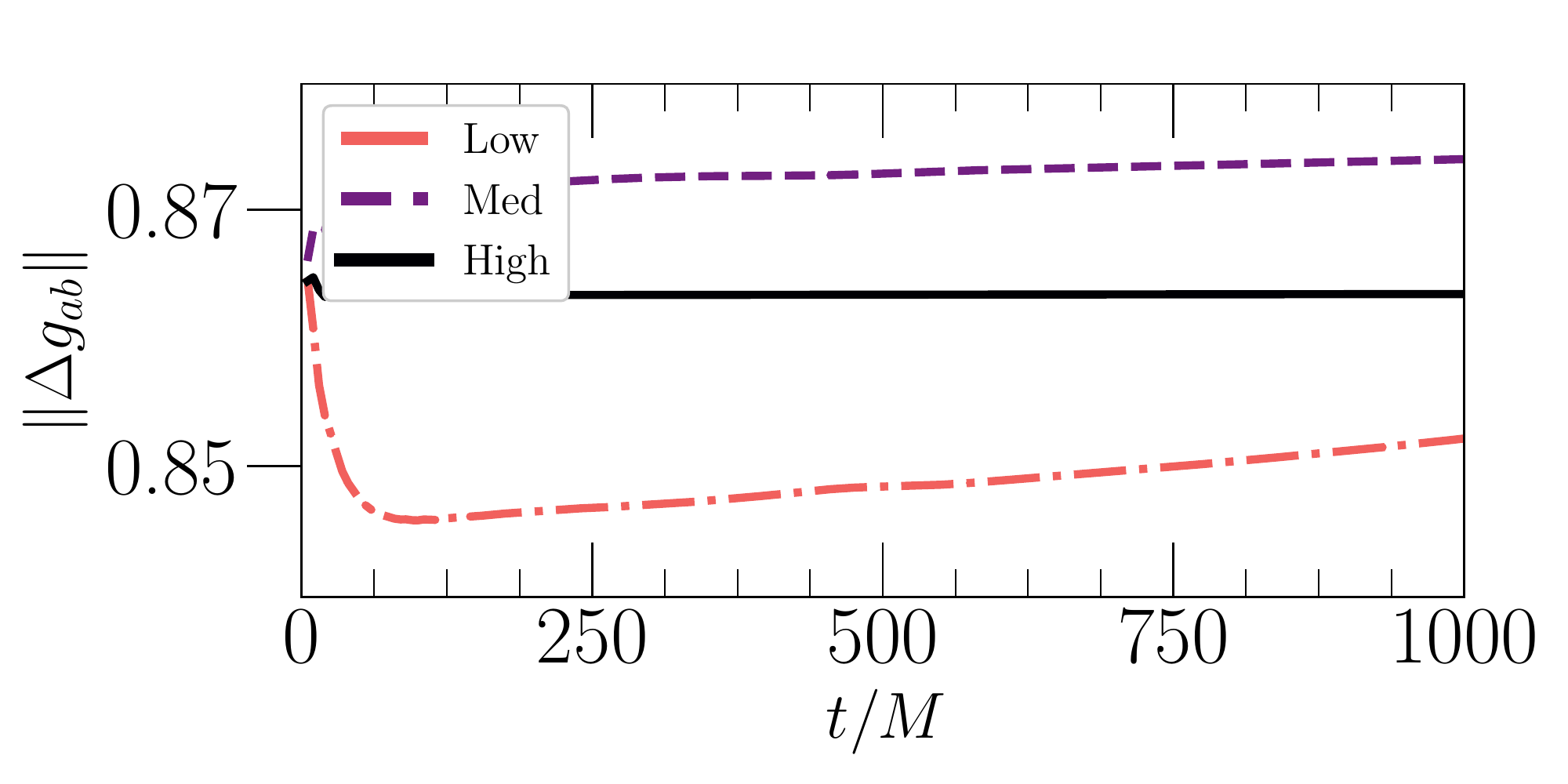}
  \caption{Metric perturbation $\Delta g_{ab}$ on a Kerr background with $\chi = 0.1$. We present the behavior at low, medium, and high resolutions, and find that we increase the numerical resolution, the level of linear growth in time decreases.}
\end{figure}

In Fig.~\ref{fig:Spin0p1EvVars}, we present the behavior of the norm of the metric perturbation with time for $\chi = 0.1$ for low, medium, and high resolution. We see that as we increase resolution, $\Delta g_{ab}$ becomes more constant in time. Note that the specific value of $\| \Delta g_{ab} \|$ ($\sim 0.86$ in Fig.~\ref{fig:Spin0p1EvVars}) should be a function of $\chi$, the spin of the back hole. However, though expressions for this functional dependence exist in the slow and rapid rotation limits~\cite{Yunes:2009hc, Konno:2009kg}, and as post-Newtonian expansions~\cite{Ayzenberg:2018jip}, no closed-form, analytical expression for the functional dependence is known. 

\begin{figure}
  \centering
  \label{fig:Spin0p6EvVars}
  \includegraphics[width=\columnwidth]{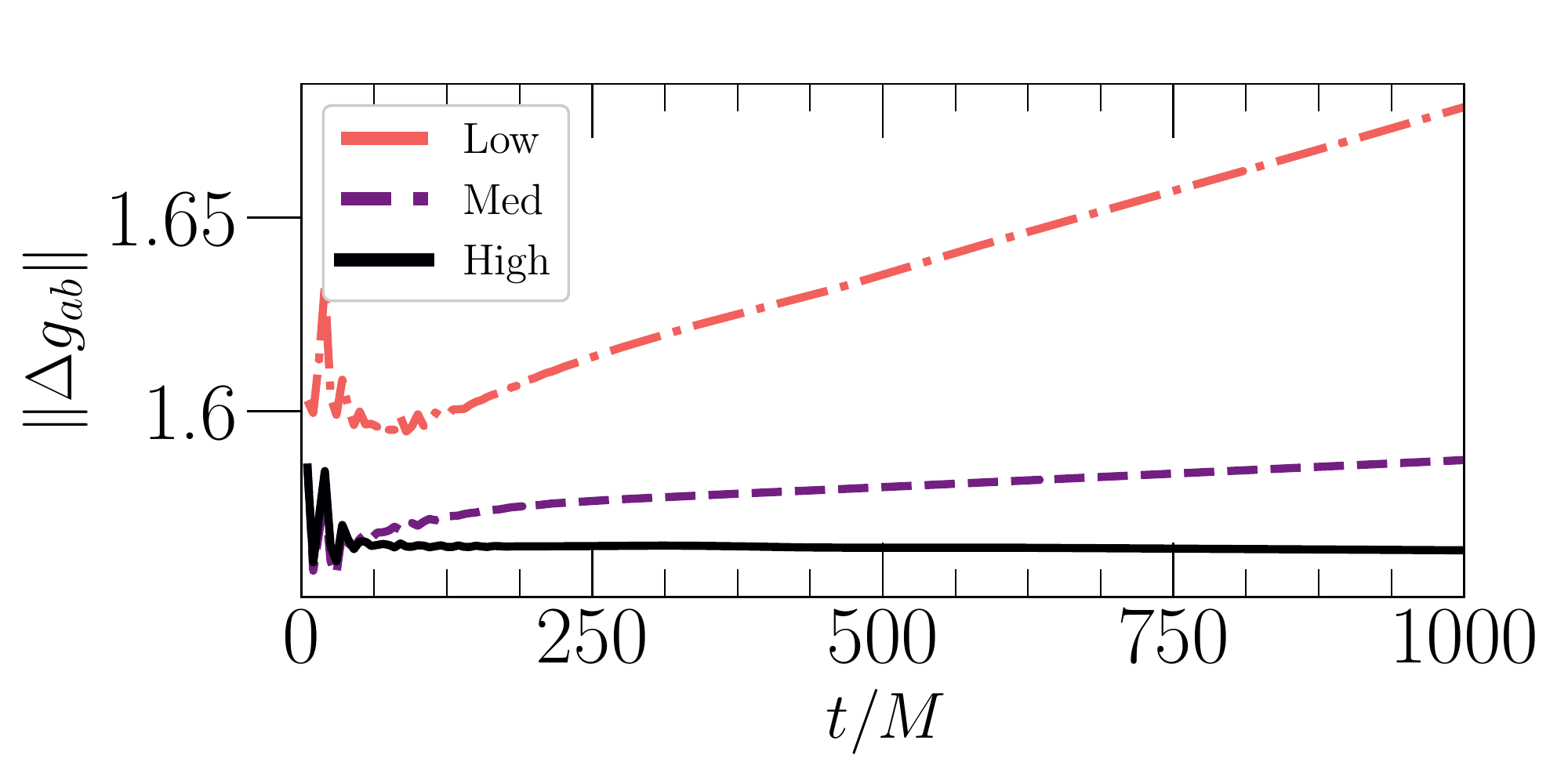}
  \caption{Similar to Fig.~\ref{fig:Spin0p1EvVars}, but for spin of $\chi = 0.6$. For each resolution, we use the initial data for $\Delta g_{ab}$ we have solved for at that resolution, and hence $\Delta g_{ab}$ has different initial values depending on resolution. We have checked that these initial values converge to the highest-resolution result. 
  }
\end{figure}

Fig.~\ref{fig:Spin0p6EvVars} similarly shows the behavior of the metric perturbation for $\chi = 0.6$. This case is particularly interesting, as it corresponds roughly to the final spin of the post-merger black holes in~\cite{Okounkova:2017yby}. We thus conclude that were we to also simulate metric perturbations in that study, we could stably evolve metric perturbations through ringdown.

\begin{figure}
  \centering
  \label{fig:SpinVars0p1}
  \includegraphics[width=\columnwidth]{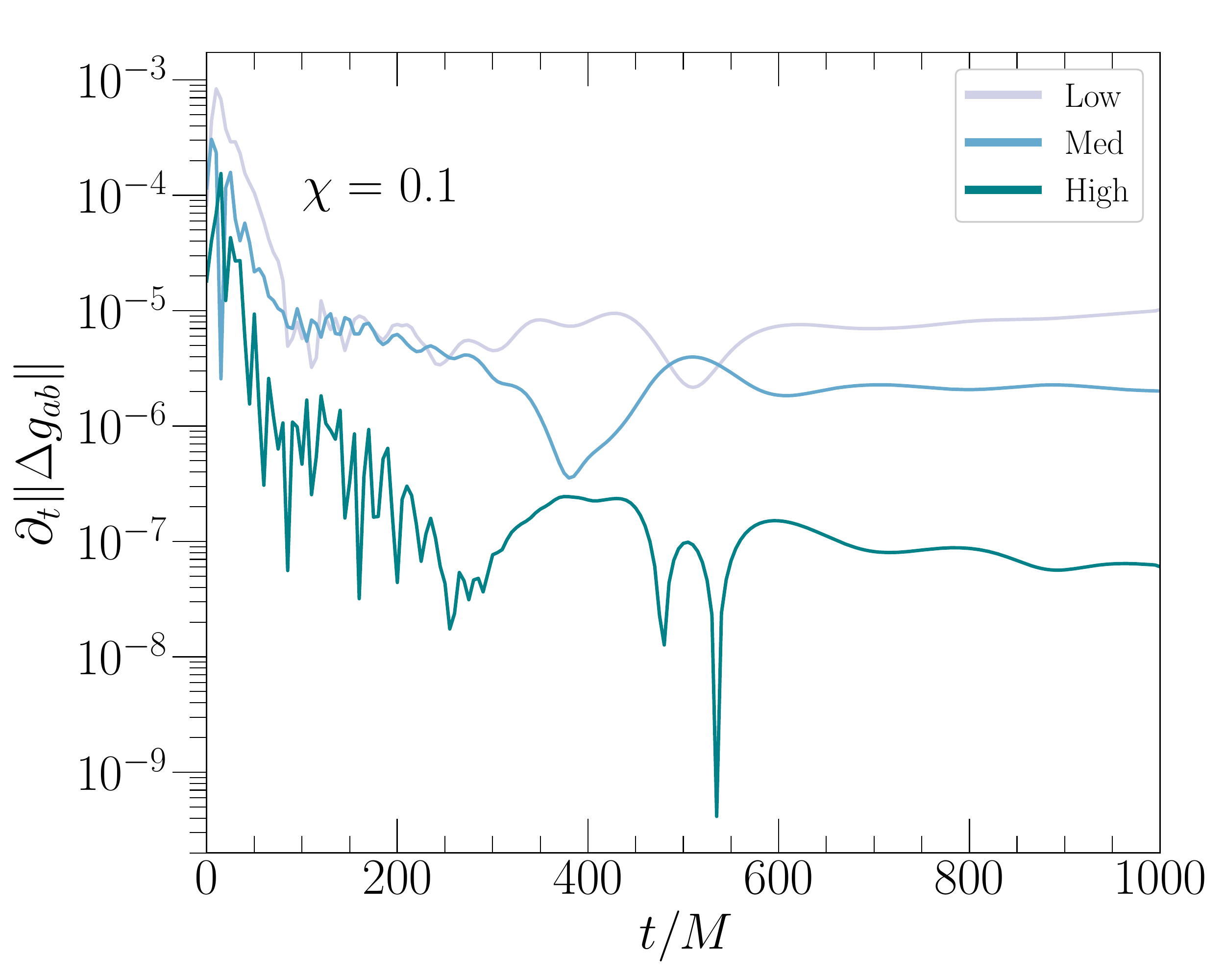}
  \caption{Behavior of the derivative of the norm of the metric perturbation with time for a background with spin $\chi = 0.1$. We plot $\pd_t \| \Delta g_{ab}\|$, the time derivative of the norm of the metric perturbation.
  Each line corresponds to a different resolution. We see that after an initial period of junk radiation, the time derivative is convergent towards zero with increasing numerical resolution.}
\end{figure}

\begin{figure}
  \centering
  \label{fig:SpinVars0p6}
  \includegraphics[width=\columnwidth]{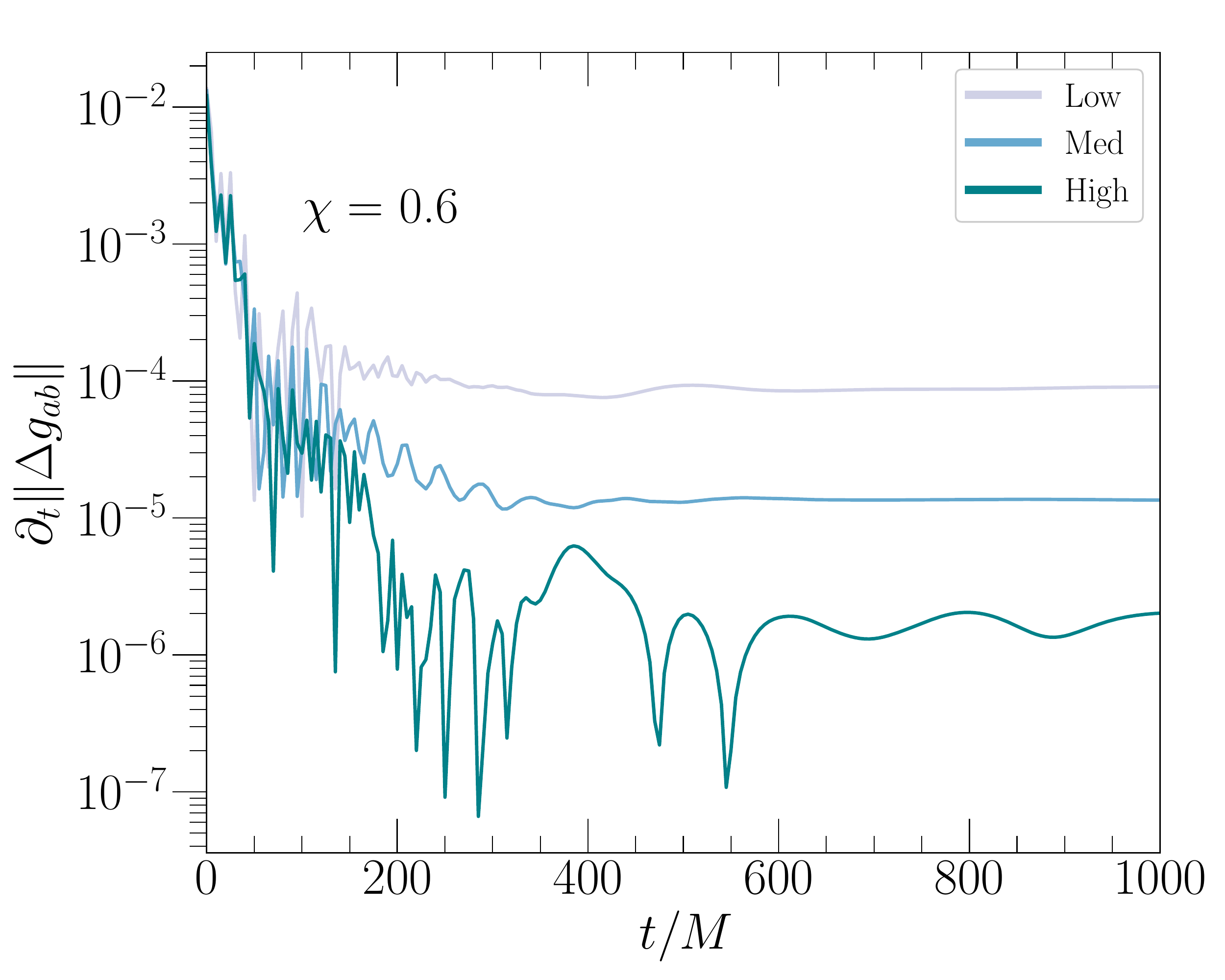}
  \caption{Similar to Fig.~\ref{fig:SpinVars0p1}, but for spin $\chi=0.6$.}
\end{figure}

\begin{figure}
  \centering
  \label{fig:SpinVars0p9}
  \includegraphics[width=\columnwidth]{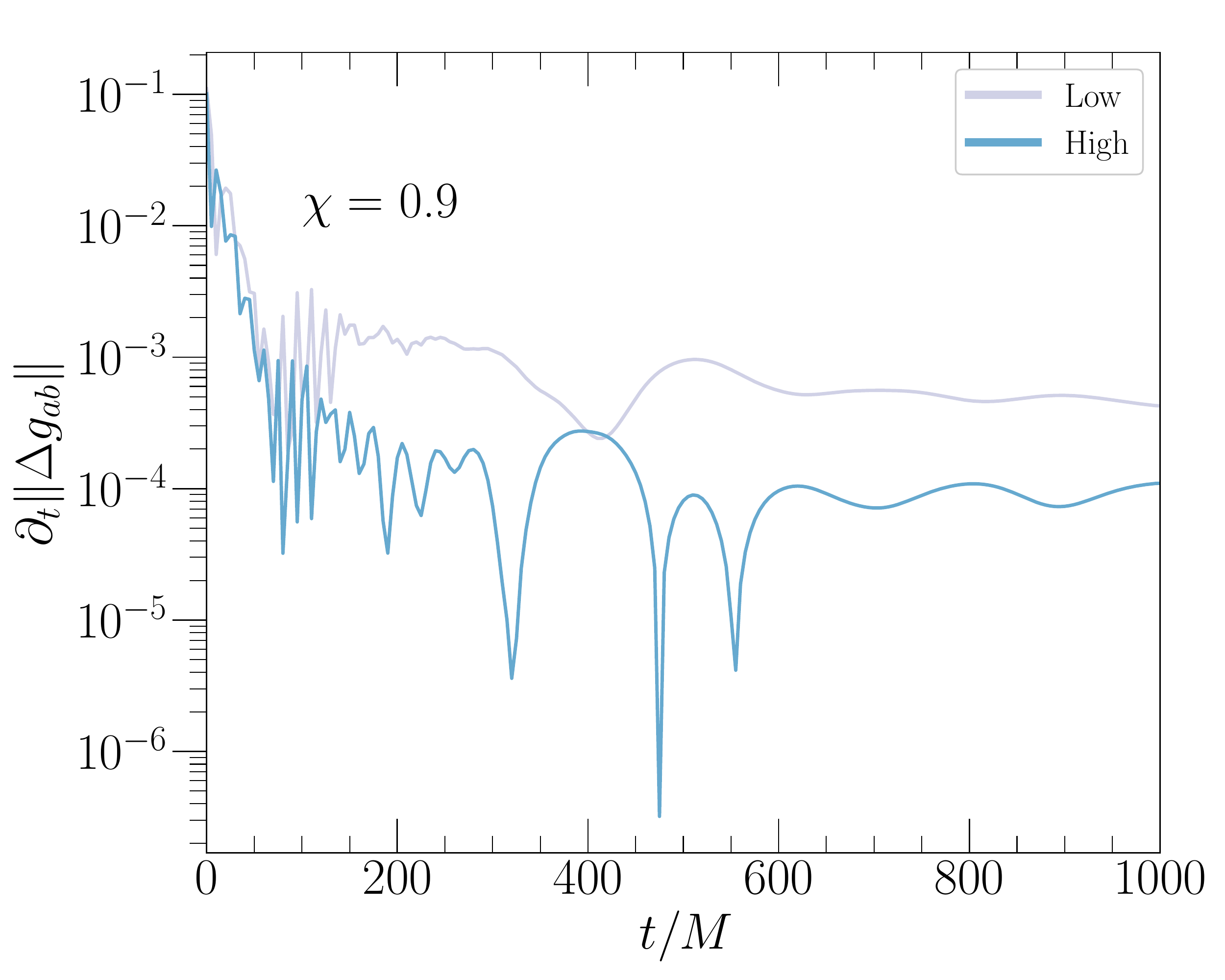}
  \caption{Similar to Fig.~\ref{fig:SpinVars0p1}, but for spin $\chi=0.9$.}
\end{figure}

For a more quantitative analysis, we show the time derivative of the norm of $\Delta g_{ab}$ in Figs.~\ref{fig:SpinVars0p1}, \ref{fig:SpinVars0p6}, and~\ref{fig:SpinVars0p9}, for $\chi=0.1$, $\chi=0.6$, and $\chi = 0.9$, for three different resolutions. Initially, there is some junk radiation (unphysical spurious radiation) present on the domain, so the first $\sim 150\,M$ (corresponding to the computational domain radius) of each figure can be ignored. 

We see that after the junk radiation has left the domain, the normalized time derivative decreases with numerical resolution, staying at a low level of $\sim 10^{-6}$ at the highest resolution\footnote{Higher spins require higher resolutions to achieve the same level of numerical accuracy in Kerr-Schild coordinates, and thus the values of the time derivatives at the same numerical resolution increase slightly with spin.}. Let us examine this result more carefully. The metric perturbation, as shown for example in Fig.~\ref{fig:Spin0p1EvVars}, exhibits linear growth in time. However, the lower numerical resolutions exhibit more linear growth than higher numerical resolutions. As shown in Fig.~\ref{fig:SpinVars0p1}, we see that with increasing numerical resolution, this linear growth converges exponentially towards zero. Thus, this linear growth is a numerical artifact, and in the limit of infinite resolution will be zero. Thus, we must evolve the metric perturbation at a high enough resolution such that the linear growth is small enough for our purposes.

How long do we need to evolve $\Delta g_{ab}$ to be confident in the stability of the field? Practically, NR gravitational waveforms typically contain $100 - 200\,M$ of ringdown signal~\cite{Mroue:2013xna}, as did the simulations we performed in~\cite{Okounkova:2017yby}. Thus, we certainly require stability on timescales of $\mathcal{O}(100)\,M$. Binary black hole simulation initial data is comprised of a{n approximate} superposition of two black hole metrics~\cite{Lovelace2009}. Thus, in the early inspiral, the spacetime is similar to that of two black holes, with a dCS metric perturbation isolated around each black hole. While binary black hole simulations typically start $\sim 5,000$ to $10,000\,M$ before merger (cf.~\cite{Mroue:2013xna}), at some point in the inspiral, strong-field dynamics take over and the spacetime is no longer a superposition of two Kerr black holes. Thus, we are interested in timescales of $\mathcal{O}(1000)\,M$, to be able to simulate the early inspiral. For one resolution, we have also evolved $\Delta g_{ab}$ on a $\chi = 0.1$ background for $10,000\,M$ (but only 1000M of evolution is shown in Fig.~\ref{fig:SpinVars0p1}). We find that the metric perturbation exhibits similar behavior on these timescales (the time derivative of the perturbed metric, $\pd_t \Delta g_{ab}$, remains at a constant level for at least $10,000\,M$).

\begin{figure}
  \centering
  \label{fig:SpinVars0p1HiResID}
  \includegraphics[width=\columnwidth]{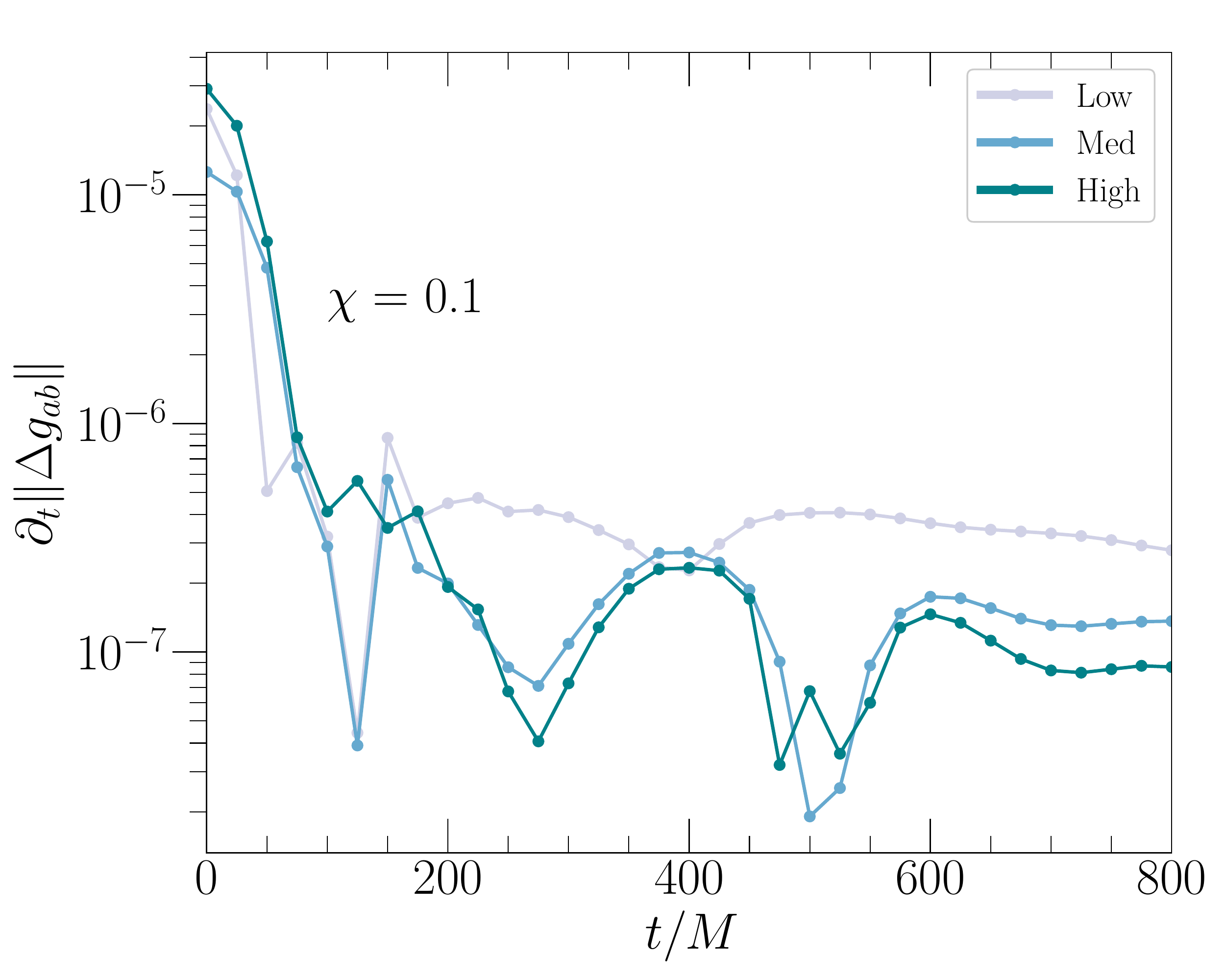}
  \caption{The structure of this figure is similar to that of  Fig.~\ref{fig:SpinVars0p1}. However, in this case, we solve for the initial data for $\Delta g_{ab}$ purely at the `High' resolution. We interpolate this data onto the `Low' and `Medium' resolution grids to give initial data at these resolutions. We see that as the simulation progresses, the linear growth in $\Delta g_{ab}$ remains at roughly the same level for all resolutions. This suggests that the zero-frequency mode in $\Delta g_{ab}$ is present in and due to the resolution of the initial data, rather than spontaneously appearing during the evolution.}
\end{figure}

Let us now discuss the origin of the linearly growing mode (a zero-frequency mode). One possibility is that it is present in the initial data for the metric perturbation, as it is in the spectrum of the differential operator. For the simulations shown in Figs.~\ref{fig:SpinVars0p1}, \ref{fig:SpinVars0p6}, and~\ref{fig:SpinVars0p9}, the evolution for each numerical resolution has its own initial data, which is solved for independently on a grid of that resolution. Thus, if the presence of the mode is purely due to the initial data, we would expect different resolutions to display various levels of linear growth, which we indeed see.  To further test this hypothesis, we can instead solve for initial data for $\Delta g_{ab}$ only at the highest resolution, and \textit{interpolate} this onto the lower-resolution grids to use for the  evolution. In Fig.~\ref{fig:SpinVars0p1HiResID}, we show the results of this procedure. We see that all three resolutions have roughly the same amount of linear growth, suggesting that the zero-frequency mode is seeded by the initial data, rather than spontaneously appearing during the evolution. Note that the growth is at the level of the highest resolution, which is still finite, and hence the growth is non-zero. This in turn tells us that in order to achieve the requisite level of numerical stability, we can use higher-resolution initial data, and perform our simulations at lower resolutions.

\clearpage

\section{Results and discussion}
\label{sec:Conclusion}

In this paper, we have aimed to test the stability of rotating black holes in dCS gravity to leading order. We have worked in order-reduced dCS, in which we perturb the dCS scalar field and metric around a GR background. We have evolved the leading-order dCS metric perturbation, sourced by the leading-order dCS scalar field coupled to the spacetime curvature of the GR background (Sec.~\ref{sec:dCSEvolution}). We  used a fully general, first-order, constraint-damping metric perturbation evolution scheme based on the generalized harmonic formalism of general relativity (Sec.~\ref{sec:Evolution}). We found that the dCS metric perturbation exhibits linear growth in time, but that the level of linear growth converges towards zero with increasing numerical resolution. 

The linear stability analysis presented in this paper shows that black holes in dCS gravity are numerically stable to leading-order perturbations in the metric. The leading-order (first non-vanishing) metric perturbation in dCS gravity occurs at second order, and thus the linear stability presented corresponds to stability at second order in the dCS order-reduction scheme. Previous studies have explored the question of black hole stability in dCS gravity~\cite{Molina:2010fb, Garfinkle:2010zx,Berti:2015itd}, but this is the first study to explore the behavior of metric perturbations on a spinning background with non-zero source. 

Linear theory has no scale, and thus the results presented in this paper can be applied to any coupling parameter $\varepsilon^2$ such that, to second order, the dCS metric is $g_{ab} + \eps^2 \Delta g_{ab}$. However, for the perturbative scheme to be valid, we must choose $\eps^2$ such that $\|\eps^2 \Delta g_{ab}\| \lesssim \|g_{ab}\|$ (cf.~\cite{Stein:2014xba} and~\cite{MashaIDPaper} for a quantitative analysis of allowed values of $\eps^2$). 

The stability of our simulations makes us confident that we can evolve dCS  metric perturbations in a binary black hole spacetime without numerical instabilities. We can use a superposition of the dCS scalar field initial data given in~\cite{Stein:2014xba} and the dCS metric perturbation initial data formalism and code used in~\cite{MashaIDPaper} to generate initial data for scalar field $\Psi$ and perturbed metric variables $\Delta g_{ab}$ and $\Delta \kappa_{abc}$. We can then evolve the scalar field as we previously have in~\cite{Okounkova:2017yby} and use this $\Psi(t)$ to source the evolution of $\Delta g_{ab}$. While we have used a stationary gauge as determined by $\Delta H_a = \Delta \Gamma_a (t = 0)$ in this work, we also have the option of rolling into a perturbed damped harmonic gauge during the binary evolution (cf.~\cite{Szilagyi:2014fna}).


\section*{Acknowledgements}

We would like to thank Leo Stein for helpful discussions,  providing us with the code used to generate the initial data in~\cite{Stein:2014xba}, and careful reading of this manuscript. We would also like to thank Luis Lehner for suggesting this project. This work was supported in part by the Sherman Fairchild Foundation, and NSF grants PHY-1708212 and PHY-1708213 at Caltech and PHY-1606654 at Cornell. Computations were performed using the Spectral Einstein Code~\cite{SpECwebsite}. All computations were performed on the Wheeler cluster at Caltech, which is supported by the Sherman Fairchild Foundation and by Caltech. 


\appendix
\section{Perturbed 2-index constraint}
\label{sec:2con}

In this appendix we derive perturbations to the generalized harmonic constraint $C_{ab}$. This constraint corresponds to a combination of the Hamiltonian and momentum constraints, and includes terms proportional to the constraint $C_{iab}$ (cf. Eq.~\eqref{eq:3con}) that are added in order to simplify the evolution equations for the constraints~\cite{Lindblom2006}. The constraint $C_{ab}$ is defined in Eqs.~43 and~44 of~\cite{Lindblom2006}, in which the time components $C_{0a}$ are called $\mathcal{F}_a$.  The expressions in~\cite{Lindblom2006} do not contain stress-energy source terms, but we include these terms here. In particular,
\begin{align}
C_{0a} &\equiv \mathcal{F}_a - 2n^b S_{ba} + n_a S_{bc} g^{bc}\,,
\end{align}
where $\mathcal{F}_a$ is the expression from~\cite{Lindblom2006}.

In terms of the variable $\kappa_{abc}$, the spatial part of the 2-index constraint is
\begin{align}
\label{eq:2conSpatial}
C_{ia} &\equiv \gamma^{jk} \pd_j \kappa_{ika} - \dfrac{1}{2}\gamma^{j}{}_a g^{cd} \pd_j \kappa_{icd} + n^b \pd_i \kappa_{0ba} \\
\nn &\quad -\frac{1}{2} n_a g^{cd} \pd_i \kappa_{0cd} + \pd_i H_a + \frac{1}{2}g^j{}_a \kappa_{jcd}\kappa_{ief} g^{ce} g^{df} \\
\nn &\quad + \frac{1}{2} \gamma^{jk} \kappa_{jcd} \kappa_{ike} g^{cd} n^e n_a \\
\nn &\quad -\gamma^{jk} \gamma^{mn} \kappa_{jma} \kappa_{ikn} \\
\nn &\quad+ \frac{1}{2} \kappa_{icd} \kappa_{0be} n_a \left(g^{cb}g^{de} + \frac{1}{2} g^{be} n^c n^d\right) \\
\nn &\quad -\kappa_{icd}\kappa_{0ba} n^c\left(g^{bd} + \frac{1}{2} n^b n^d\right) \\
\nn& \quad + \frac{1}{2} \gamma_2 \left(n_a g^{cd} - 2\delta^c_a n^d\right) C_{icd}\,,
\end{align}
and the time part is the lengthy expression

\begin{align}
\label{eq:2conTime}
&C_{0a} \equiv \quad- 2n^b S_{ba} + n_a S_{bc} g^{bc} \\
\nn &\quad+ \frac{1}{2} g^i_a g^{bc} \pd_i \kappa_{0bc} - \gamma^{ij} \pd_i \kappa_{0ja} - \gamma^{ij} n^b \pd_i \kappa_{jba}\\
\nn&\quad + \frac{1}{2} n_a g^{bc} \gamma^{ij} \pd_i \kappa_{jbc} + n_a \gamma^{ij} \pd_i H_j \\
\nn&\quad + g^i_a \kappa_{ijb}\gamma^{jk}\kappa_{kcd}\Big(g^{bd}n^c - \frac{1}{2} g^{cd}n^b\Big) \\
\nn&\quad -g^i_a n^b \pd_i H_b + \gamma^{ij} \kappa_{icd} \kappa_{jba} g^{bc} n^d \\
\nn&\quad - \frac{1}{2} n_a \gamma^{ij}\gamma^{mn}\kappa_{imc} \kappa_{njd} g^{cd} \\
\nn&\quad - \frac{1}{4} n_a \gamma^{ij} \kappa_{icd} \kappa_{jbe} g^{cb} g^{de} + \frac{1}{4} n_a \kappa_{0cd} \kappa_{0be}g^{cb}g^{de} \\
\nn&\quad -\gamma^{ij}H_i \kappa_{0ja} - n^b \gamma^{ij} \kappa_{0bi} \kappa_{0ja} \\
\nn&\quad -\frac{1}{4}g^i_a \kappa_{icd}n^c n^d \kappa_{0be} g^{be} + \frac{1}{2} n_a \kappa_{0cd} \kappa_{0be} g^{ce} n^d n^b \\
\nn&\quad + g^i_a \kappa_{icd} \kappa_{0be} n^c n^b g^{de} - \gamma^{ij} \kappa_{iba} n^b \kappa_{0je} n^e \\
\nn&\quad - \frac{1}{2} \gamma^{ij} \kappa_{icd} n^c n^d \kappa_{0ja} - \gamma^{ij} H_i \kappa_{jba} n^b \\
\nn&\quad + g^i_a \kappa_{icd} H_b g^{bc} n^d \\
\nn&\quad + \gamma_2 \Big(\gamma^{id}C_{ida} - \frac{1}{2} g^i_a g^{cd}C_{icd}\Big) \\
\nn&\quad + \frac{1}{2} n_a \kappa_{0cd} g^{cd} H_b n^b - n_a \gamma^{ij} \kappa_{ijc} H_d g^{cd} \\
\nn&\quad + \frac{1}{2} n_a \gamma^{ij} H_i \kappa_{jcd} g^{cd}\,.
\end{align}

Perturbing Eq.~\eqref{eq:2conSpatial} to obtain the perturbation to the spatial part of the 2-index constraint, we find
\begin{widetext}
\small{
\begin{align}
\label{eq:h2conSpatial}
&\Delta C_{ia} \equiv \Delta \gamma^{jk} \pd_j \kappa_{ika} + \gamma^{jk} \pd_j \Delta \kappa_{ika}  - \dfrac{1}{2}\Delta \gamma^{j}{}_a g^{cd} \pd_j \kappa_{icd} - \dfrac{1}{2}\gamma^{j}{}_a \Delta g^{cd} \pd_j \kappa_{icd}  \\
\nn &\quad - \dfrac{1}{2}\gamma^{j}{}_a g^{cd} \pd_j \Delta \kappa_{icd} + \Delta n^b \pd_i \kappa_{0ba} + n^b \pd_i \Delta \kappa_{0ba}  -\frac{1}{2} \Delta n_a g^{cd} \pd_i \kappa_{0cd} -\frac{1}{2} n_a \Delta g^{cd} \pd_i \kappa_{0cd} \\
\nn &\quad -\frac{1}{2} n_a g^{cd} \pd_i \Delta \kappa_{0cd} + \pd_i \Delta H_a + \frac{1}{2}\Delta g^j{}_a \kappa_{jcd}\kappa_{ief} g^{ce} g^{df} + \frac{1}{2}g^j{}_a \Delta \kappa_{jcd}\kappa_{ief} g^{ce} g^{df} \\
\nn &\quad + \frac{1}{2}g^j{}_a \kappa_{jcd}\Delta \kappa_{ief} g^{ce} g^{df} + \frac{1}{2}g^j{}_a \kappa_{jcd}\kappa_{ief} \Delta g^{ce} g^{df} + \frac{1}{2}g^j{}_a \kappa_{jcd}\kappa_{ief} g^{ce} \Delta g^{df} + \frac{1}{2} \Delta \gamma^{jk} \kappa_{jcd} \kappa_{ike} g^{cd} n^e n_a \\
\nn &\quad + \frac{1}{2} \gamma^{jk} \Delta \kappa_{jcd} \kappa_{ike} g^{cd} n^e n_a + \frac{1}{2} \gamma^{jk} \kappa_{jcd} \Delta \kappa_{ike} g^{cd} n^e n_a  + \frac{1}{2} \gamma^{jk} \kappa_{jcd} \kappa_{ike} \Delta g^{cd} n^e n_a + \frac{1}{2} \gamma^{jk} \kappa_{jcd} \kappa_{ike} g^{cd} \Delta n^e n_a \\
\nn&\quad + \frac{1}{2} \gamma^{jk} \kappa_{jcd} \kappa_{ike} g^{cd} n^e \Delta n_a -\Delta \gamma^{jk} \gamma^{mn} \kappa_{jma} \kappa_{ikn}-\gamma^{jk} \Delta \gamma^{mn} \kappa_{jma} \kappa_{ikn} -\gamma^{jk} \gamma^{mn} \Delta \kappa_{jma} \kappa_{ikn} -\gamma^{jk} \gamma^{mn} \kappa_{jma} \Delta \kappa_{ikn} \\
\nn &\quad+ \frac{1}{2} (\Delta \kappa_{icd} \kappa_{0be} n_a + \kappa_{icd} \Delta \kappa_{0be} n_a + \kappa_{icd} \kappa_{0be} \Delta n_a) \times  \left(g^{cb}g^{de} + \frac{1}{2} g^{be} n^c n^d\right) \\
\nn &\quad+ \frac{1}{2} \kappa_{icd} \kappa_{0be} n_a \Big(\Delta g^{cb}g^{de} + g^{cb}\Delta g^{de} + \frac{1}{2} (\Delta g^{be} n^c n^d + g^{be} \Delta n^c n^d + g^{be} n^c \Delta n^d )\Big) \\
\nn &\quad -(\Delta \kappa_{icd}\kappa_{0ba} n^c + \kappa_{icd}\Delta \kappa_{0ba} n^c + \kappa_{icd}\kappa_{0ba} \Delta n^c ) \times \Big(g^{bd} + \frac{1}{2} n^b n^d \Big) \\
\nn &\quad -\kappa_{icd}\kappa_{0ba} n^c \Big(\Delta g^{bd} + \frac{1}{2} \Delta n^b n^d +  \frac{1}{2} n^b \Delta n^d \Big) \\
\nn& \quad + \frac{1}{2} \gamma_2 \left(\Delta n_a g^{cd} + n_a \Delta g^{cd} - 2\delta^c_a \Delta n^d\right) C_{icd}+ \frac{1}{2} \gamma_2 \left(n_a g^{cd} - 2\delta^c_a n^d\right) \Delta C_{icd} \,,
\end{align}
}
\end{widetext}
where $\Delta C_{icd}$ is the perturbed 3-index constraint as defined in Eq.~\eqref{eq:h3con}.

Finally, the perturbation to the time part of the 2-index constraint is
\begin{widetext}
\small{
\begin{align}
\label{eq:h2conTime}
&\Delta C_{0a} \equiv -2\Delta n^b S_{ba} - 2n^b \Delta S_{ba}  +  \Delta n_a S_{bc} g^{bc} + n_a \Delta S_{bc} g^{bc} + n_a S_{bc} \Delta g^{bc} \\
\nn &\quad+ \frac{1}{2} \Big( \Delta g^i_a g^{bc} \pd_i \kappa_{0bc} + g^i_a \Delta g^{bc} \pd_i \kappa_{0bc} + g^i_a g^{bc} \pd_i \Delta \Big) \kappa_{0bc} - \Delta \gamma^{ij} \pd_i \kappa_{0ja} - \gamma^{ij} \pd_i \Delta \kappa_{0ja} \\
\nn&\quad - \Delta \gamma^{ij} n^b \pd_i \kappa_{jba} - \gamma^{ij} \Delta n^b \pd_i \kappa_{jba} - \gamma^{ij} n^b \pd_i \Delta \kappa_{jba}\\
\nn&\quad + \frac{1}{2} \Big( \Delta n_a g^{bc} \gamma^{ij} \pd_i \kappa_{jbc} + n_a \Delta g^{bc} \gamma^{ij} \pd_i \kappa_{jbc} + n_a g^{bc} \Delta \gamma^{ij} \pd_i \kappa_{jbc} + n_a g^{bc} \gamma^{ij} \pd_i \Delta \kappa_{jbc}\Big) \\
\nn&\quad + \Delta n_a \gamma^{ij} \pd_i H_j + n_a \Delta \gamma^{ij} \pd_i H_j + n_a \gamma^{ij} \pd_i \Delta H_j \\
\nn&\quad + \Big(\Delta g^i_a \kappa_{ijb}\gamma^{jk}\kappa_{kcd} + g^i_a \Delta \kappa_{ijb}\gamma^{jk}\kappa_{kcd} + g^i_a \kappa_{ijb}\Delta \gamma^{jk}\kappa_{kcd} + g^i_a \kappa_{ijb}\gamma^{jk}\Delta \kappa_{kcd}\Big) \times \Big(g^{bd}n^c - \frac{1}{2} g^{cd}n^b\Big) \\
\nn&\quad + g^i_a \kappa_{ijb}\gamma^{jk}\kappa_{kcd}\Big(\Delta g^{bd}n^c + g^{bd}\Delta n^c - \frac{1}{2} \Delta g^{cd}n^b  - \frac{1}{2} g^{cd}\Delta n^b\Big) \\
\nn&\quad -\Delta g^i_a n^b \pd_i H_b -g^i_a \Delta n^b \pd_i H_b  -g^i_a n^b \pd_i \Delta H_b \\
\nn&\quad + \Delta \gamma^{ij} \kappa_{icd} \kappa_{jba} g^{bc} n^d + \gamma^{ij} \Delta \kappa_{icd} \kappa_{jba} g^{bc} n^d + \gamma^{ij} \kappa_{icd} \Delta \kappa_{jba} g^{bc} n^d + \gamma^{ij} \kappa_{icd} \kappa_{jba} \Delta g^{bc} n^d + \gamma^{ij} \kappa_{icd} \kappa_{jba} g^{bc} \Delta n^d \\
\nn&\quad - \frac{1}{2} \Big(\Delta n_a \gamma^{ij}\gamma^{mn}\kappa_{imc} \kappa_{njd} g^{cd} + n_a \Delta \gamma^{ij}\gamma^{mn}\kappa_{imc} \kappa_{njd} g^{cd} + n_a \gamma^{ij}\Delta \gamma^{mn}\kappa_{imc} \kappa_{njd} g^{cd} \\
\nn &\quad \quad + n_a \gamma^{ij}\gamma^{mn}\Delta \kappa_{imc} \kappa_{njd} g^{cd} + n_a \gamma^{ij}\gamma^{mn}\kappa_{imc} \Delta \kappa_{njd} g^{cd} + n_a \gamma^{ij}\gamma^{mn}\kappa_{imc} \kappa_{njd} \Delta g^{cd}\Big) \\
\nn&\quad - \frac{1}{4} \Big( \Delta n_a \gamma^{ij} \kappa_{icd} \kappa_{jbe} g^{cb} g^{de} + n_a \Delta \gamma^{ij} \kappa_{icd} \kappa_{jbe} g^{cb} g^{de} + n_a \gamma^{ij} \Delta \kappa_{icd} \kappa_{jbe} g^{cb} g^{de} \\
\nn& \quad \quad + n_a \gamma^{ij} \kappa_{icd} \Delta \kappa_{jbe} g^{cb} g^{de} + n_a \gamma^{ij} \kappa_{icd} \kappa_{jbe} \Delta g^{cb} g^{de} + n_a \gamma^{ij} \kappa_{icd} \kappa_{jbe} g^{cb} \Delta g^{de} \Big)
\\ \nn &\quad + \frac{1}{4} \Big( \Delta n_a \kappa_{0cd} \kappa_{0be}g^{cb}g^{de} + n_a \Delta \kappa_{0cd} \kappa_{0be}g^{cb}g^{de} + n_a \kappa_{0cd} \Delta \kappa_{0be}g^{cb}g^{de} + n_a \kappa_{0cd} \kappa_{0be}\Delta g^{cb}g^{de} + n_a \kappa_{0cd} \kappa_{0be}g^{cb}\Delta g^{de}\Big) \\
\nn&\quad -\Delta \gamma^{ij}H_i \kappa_{0ja} -\gamma^{ij}\Delta H_i \kappa_{0ja} -\gamma^{ij}H_i \Delta \kappa_{0ja} -\Delta n^b \gamma^{ij} \kappa_{0bi} \kappa_{0ja} - n^b \Delta \gamma^{ij} \kappa_{0bi} \kappa_{0ja} - n^b \gamma^{ij} \Delta \kappa_{0bi} \kappa_{0ja} - n^b \gamma^{ij} \kappa_{0bi} \Delta \kappa_{0ja} \\
\nn&\quad -\frac{1}{4}\Big(\Delta g^i_a \kappa_{icd}n^c n^d \kappa_{0be} g^{be} + g^i_a \Delta \kappa_{icd}n^c n^d \kappa_{0be} g^{be} + g^i_a \kappa_{icd} \Delta n^c n^d \kappa_{0be} g^{be} + g^i_a \kappa_{icd}n^c \Delta n^d \kappa_{0be} g^{be} \\
\nn &\quad \quad + g^i_a \kappa_{icd}n^c n^d \Delta \kappa_{0be} g^{be} + g^i_a \kappa_{icd}n^c n^d \kappa_{0be} \Delta g^{be}\big) + \frac{1}{2} \Big(\Delta n_a \kappa_{0cd} \kappa_{0be} g^{ce} n^d n^b + n_a \Delta \kappa_{0cd} \kappa_{0be} g^{ce} n^d n^b \\
\nn & \quad \quad + n_a \kappa_{0cd} \Delta \kappa_{0be} g^{ce} n^d n^b + n_a \kappa_{0cd} \kappa_{0be} \Delta g^{ce} n^d n^b + n_a \kappa_{0cd} \kappa_{0be} g^{ce} \Delta n^d n^b + n_a \kappa_{0cd} \kappa_{0be} g^{ce} n^d \Delta n^b\Big) \\
\nn&\quad + \Delta g^i_a \kappa_{icd} \kappa_{0be} n^c n^b g^{de} + g^i_a \Delta \kappa_{icd} \kappa_{0be} n^c n^b g^{de} + g^i_a \kappa_{icd} \Delta \kappa_{0be} n^c n^b g^{de} + g^i_a \kappa_{icd} \kappa_{0be} \Delta n^c n^b g^{de} \\
\nn &\quad + g^i_a \kappa_{icd} \kappa_{0be} n^c \Delta n^b g^{de} + g^i_a \kappa_{icd} \kappa_{0be} n^c n^b \Delta g^{de}- \Delta \gamma^{ij} \kappa_{iba} n^b \kappa_{0je} n^e - \gamma^{ij} \Delta \kappa_{iba} n^b \kappa_{0je} n^e - \gamma^{ij} \kappa_{iba}\Delta  n^b \kappa_{0je} n^e \\
\nn &\quad - \gamma^{ij} \kappa_{iba} n^b \Delta \kappa_{0je} n^e - \gamma^{ij} \kappa_{iba} n^b \kappa_{0je} \Delta n^e - \frac{1}{2} \Big(\Delta \gamma^{ij} \kappa_{icd} n^c n^d \kappa_{0ja} + \gamma^{ij} \Delta \kappa_{icd} n^c n^d \kappa_{0ja}+ \gamma^{ij} \kappa_{icd} \Delta n^c n^d \kappa_{0ja} \\
\nn &\quad \quad + \gamma^{ij} \kappa_{icd} n^c \Delta n^d \kappa_{0ja}+ \gamma^{ij} \kappa_{icd} n^c n^d \kappa_{0ja}+ \gamma^{ij} \kappa_{icd} n^c n^d \Delta \kappa_{0ja}\Big) \\
\nn & \quad - \Delta \gamma^{ij} H_i \kappa_{jba} n^b - \gamma^{ij} \Delta H_i \kappa_{jba} n^b - \gamma^{ij} H_i \Delta \kappa_{jba} n^b - \gamma^{ij} H_i \kappa_{jba} \Delta n^b + \Delta g^i_a \kappa_{icd} H_b g^{bc} n^d + g^i_a \Delta \kappa_{icd} H_b g^{bc} n^d  \\
\nn & \quad + g^i_a \kappa_{icd} \Delta H_b g^{bc} n^d  + g^i_a \kappa_{icd} H_b \Delta g^{bc} n^d  + g^i_a \kappa_{icd} H_b g^{bc} \Delta n^d \\
\nn&\quad + \gamma_2 \Big(\Delta \gamma^{id}C_{ida} + \gamma^{id}\Delta C_{ida} - \frac{1}{2} \Big(\Delta g^i_a g^{cd}C_{icd} + g^i_a \Delta g^{cd}C_{icd} + g^i_a g^{cd}\Delta C_{icd}\Big)\Big) \\
\nn&\quad +   \frac{1}{2} \Big(\Delta n_a \kappa_{0cd} g^{cd} H_b n^b + n_a \Delta \kappa_{0cd} g^{cd} H_b n^b + n_a \kappa_{0cd} \Delta g^{cd} H_b n^b + n_a \kappa_{0cd} g^{cd} \Delta H_b n^b + n_a \kappa_{0cd} g^{cd} H_b \Delta n^b \Big) \\
\nn &\quad - \Delta n_a \gamma^{ij} \kappa_{ijc} H_d g^{cd} - n_a \Delta \gamma^{ij} \kappa_{ijc} H_d g^{cd} - n_a \gamma^{ij} \Delta \kappa_{ijc} H_d g^{cd} - n_a \gamma^{ij} \kappa_{ijc} \Delta H_d g^{cd} - n_a \gamma^{ij} \kappa_{ijc} H_d \Delta g^{cd}\\
\nn&\quad + \frac{1}{2} \Big(\Delta n_a \gamma^{ij} H_i \kappa_{jcd} g^{cd} + n_a \Delta \gamma^{ij} H_i \kappa_{jcd} g^{cd} + n_a \gamma^{ij} \Delta H_i \kappa_{jcd} g^{cd} + n_a \gamma^{ij} H_i \Delta \kappa_{jcd} g^{cd} + n_a \gamma^{ij} H_i \kappa_{jcd} \Delta g^{cd}\Big)\,,
\end{align}
}
\end{widetext}
where $\Delta S_{ab}$ is the perturbation to the source term as given by Eq.~\eqref{eq:hSab}. We combine Eqs.~\eqref{eq:h2conSpatial} and~\eqref{eq:h2conTime} into one overall constraint,
\begin{align}
\label{eq:h2con}
\Delta C_{ab} = (\Delta C_{0a}, \Delta C_{ia})\,.
\end{align}

\section{Code tests}
\label{sec:CodeTestsAppendix}

In order to have confidence in our dCS metric perturbation evolution results, we perform a suite of tests to check the accuracy of our metric perturbation evolution code. For each test, we check the convergence of the perturbed constraints derived in Sec.~\ref{sec:Constraints}. Note that the results of these tests do not contain new physics, but rather serve as a check of our implementation of the metric perturbation evolution equations (Eqs.~\eqref{eq:dtpsi},~\eqref{eq:dtkappai}, and ~\eqref{eq:dtkappa0}). 

\subsection{Multipolar wave evolution}
\label{sec:MultipolarTest}

We first evolve a multipolar wave in the transverse-traceless gauge on a flat background~\cite{PhysRevD.26.745, Rinne:2008ig}. This evolution takes place on a domain with only one (outer) boundary, where we set the boundary condition given in Eq.~\eqref{eq:BoundaryCondition}. We wish to test the numerical evolution against the analytic solution.  However, some of the terms in the evolution equations we are testing will vanish because the analytic solution has symmetries.  To remove these symmetries, we perform a coordinate transformation of the form
\begin{align}
\label{eq:CubicScale}
r \to a \bar{r} + (a_0 - a) \frac{\bar r^3}{R^2}\,,
\end{align}
where $r \equiv \sqrt{x^2 + y^2 + z^2}$ in Cartesian grid coordinates, $R$ and $a_0$ are constants, and $a(t)$ is a (time-dependent) function. We add an additional coordinate translation of the form 
\begin{align}
\label{eq:CoordShift}
\bar{x}^i \to \bar{x}^i + C^i\,,
\end{align}
for some vector $C^i$. 

\begin{figure}
  \includegraphics[width=\columnwidth]{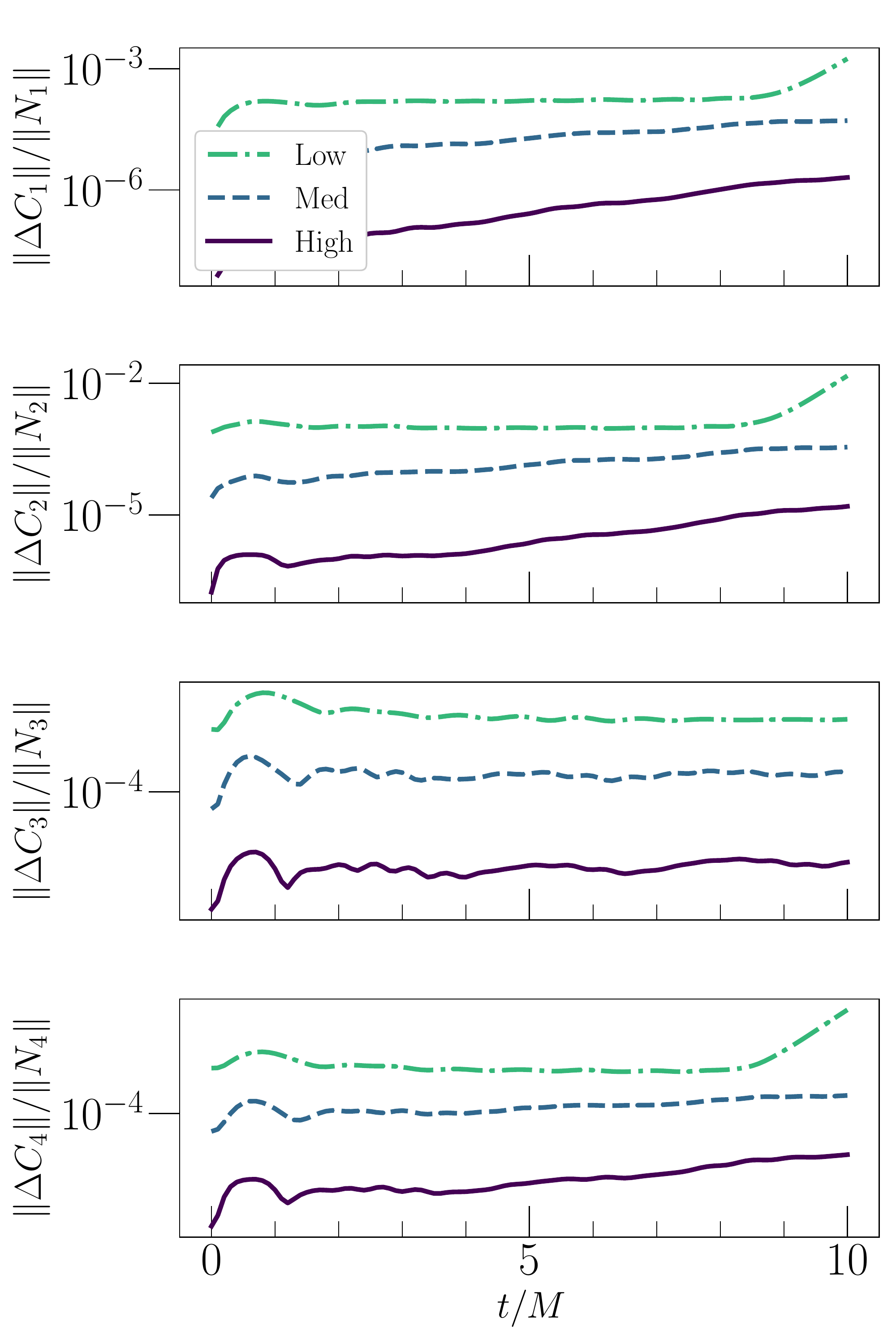}
  \caption{Constraints for evolution of a transformed multipolar wave perturbation on flat space, as described in Sec.~\ref{sec:MultipolarTest}. For each constraint $\Delta C_A$, we compute the L2 norm of the constraint over the entire computational domain ($\|\Delta C_1\|$ for the 1-index constraint, for example), and divide by the L2 norm of its normalization factor ($\| N_A \|$) (cf. Sec.~\ref{sec:Constraints}).  We see that the constraints converge exponentially with numerical resolution.  
  }
  \label{fig:MultipolarConstraints}
\end{figure}

\begin{figure}
  \centering
  \includegraphics[width=\columnwidth]{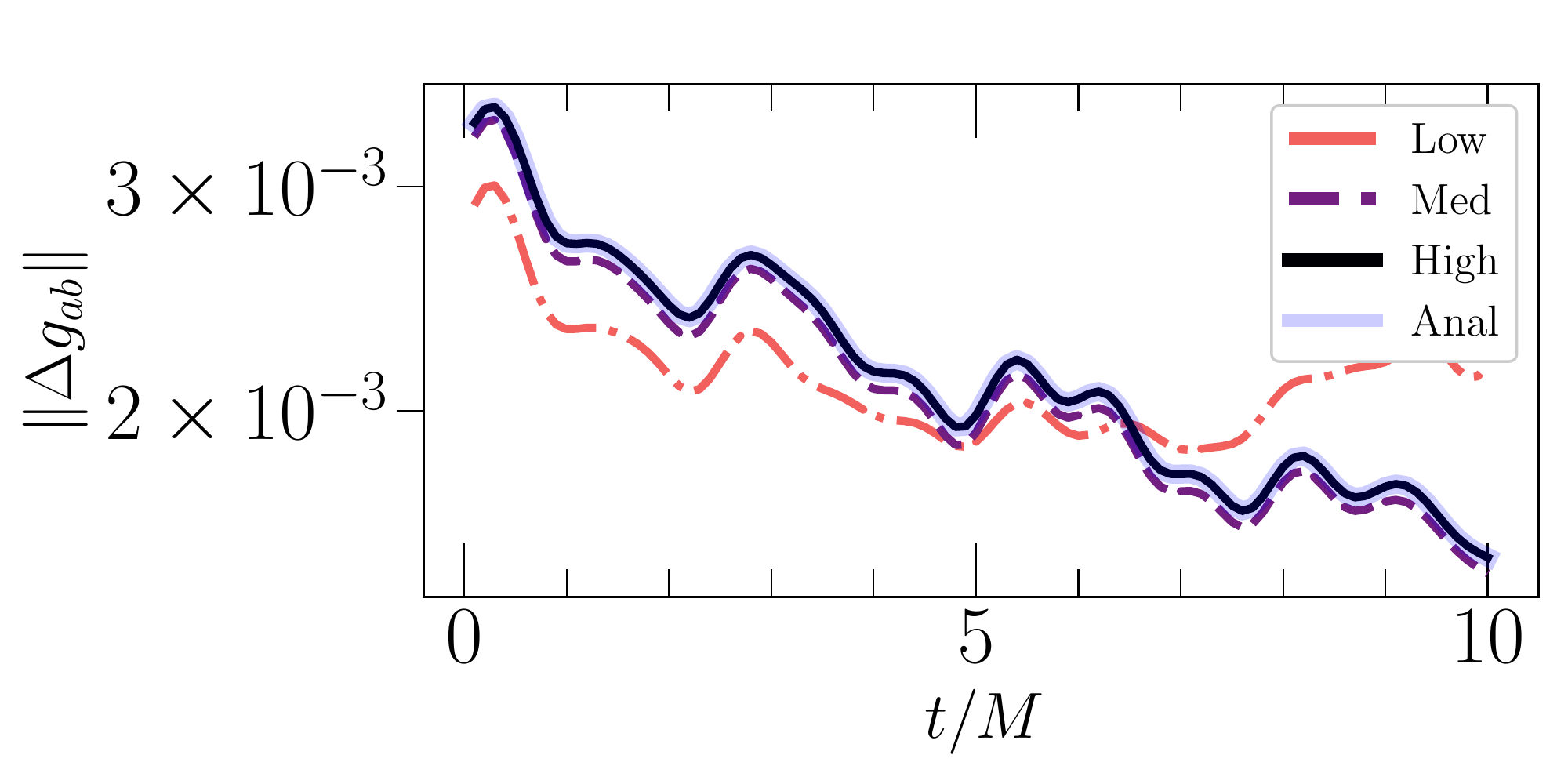}
  \caption{Behavior of $\Delta g_{ab}$ for the multipolar wave test described in Sec.~\ref{sec:MultipolarTest} for low, medium, and high resolution. We see that the the value of the metric perturbation decreases as the  wave propagates toward $R \to \infty$ (and leaves the computational domain), and that with increasing resolution the behavior of the variables converges to the highest-resolution value. We additionally plot the analytical solution for the behavior of the multipolar wave, which sits on top of the highest-resolution result. 
  }
  \label{fig:MultipolarVars}
\end{figure}

We evolve an outgoing $l = 2, m = 2$ multipolar wave. This has a Gaussian profile, with an initial width of $1\,M$, amplitude of $0.01$, and center of $10\,M$. For the transformations given in Eqs.~\eqref{eq:CubicScale} and~\eqref{eq:CoordShift}, we choose $R = 40\,M$, $a_0 = 1.3$, $a(t) = 1 + 0.001 t^2/M^2$ and $C^i = (2.0, -4.0, 3.0)\,M$. We evolve on a grid of nested spherical shells around a filled sphere, with an outer boundary of $R = 35\,M$. Each shell has 8 radial spectral basis functions and 4 angular spectral basis functions at the lowest resolution, with 4 more basis functions added in each direction as we increase resolution. We find that the perturbed constraints, shown in Fig.~\ref{fig:MultipolarConstraints}, converge exponentially, and that the perturbed variables shown in Fig.~\ref{fig:MultipolarVars} evolve toward zero (as the data leaves the domain) in a convergent way. Additionally, we check that our results converge to the known analytic solution.

\clearpage

\subsection{Small data on Schwarzschild}
\label{sec:SmallData}

\begin{figure}
  \centering
  \includegraphics[width=\columnwidth]{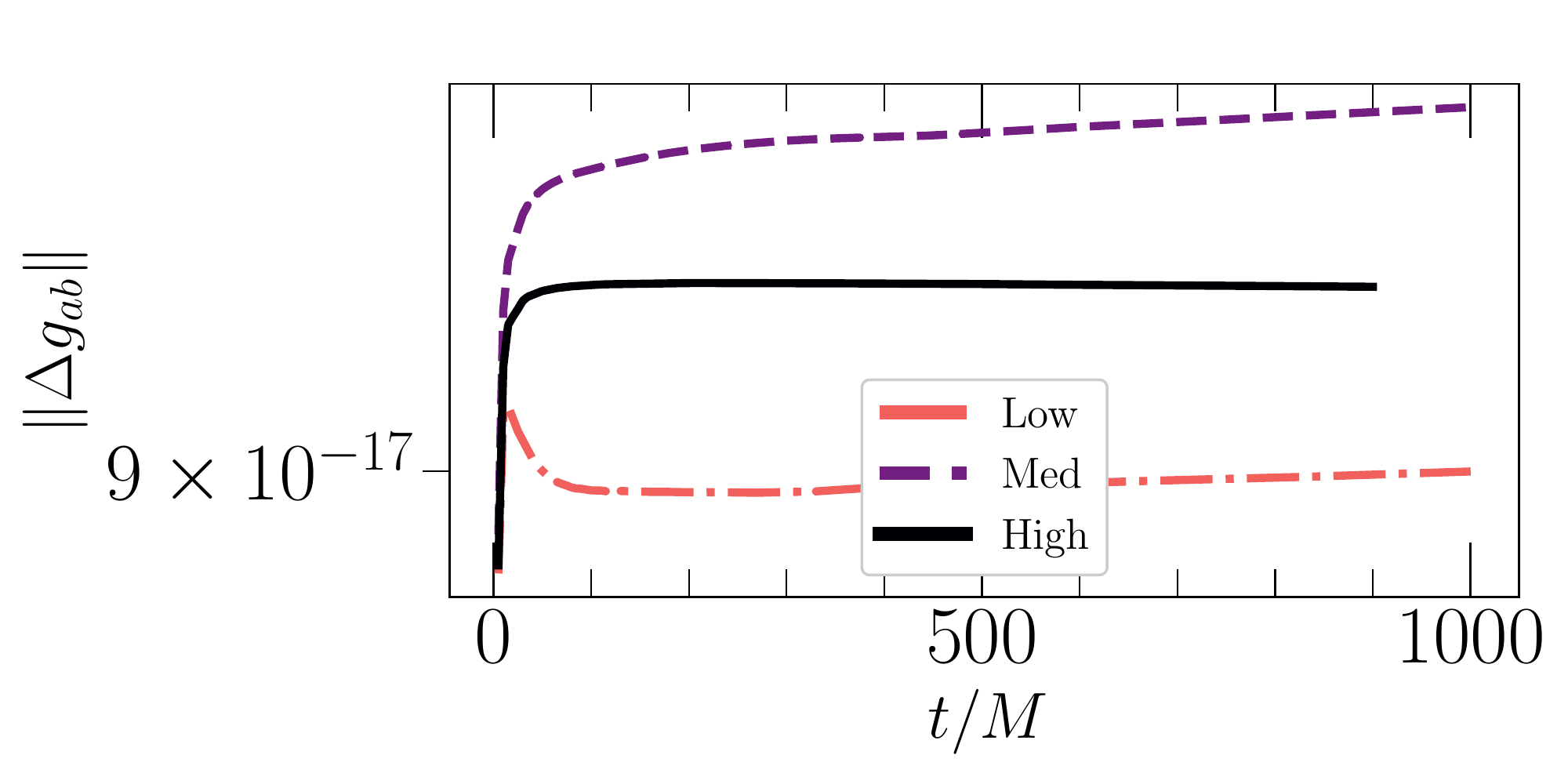}
  \caption{Behavior of $\Delta g_{ab}$ for the small data on Schwarzschild test described in Sec.~\ref{sec:SmallData}. We see that with increasing time, the field with initial magnitude of $\sim 10^{-16}$ remains close to roundoff error.
  }
  \label{fig:DiabolicalVars}
\end{figure}

We perform a test where we initially set each component of $\Delta g_{ab}$ to be a different number close to machine precision $(10^{-16})$ at each point on the domain, thus seeding any instabilities that might be present. We apply filtering to the spectral scheme in order to minimize the growth of high-frequency modes~\cite{Szilagyi:2009qz}, and choose damping parameters $\gamma_0$ and $\gamma_2$ to be larger close to the horizon. We check that as the evolution progresses, the constraints and the values of $\Delta g_{ab}$ and $\Delta \kappa_{abc}$ remain close to numerical truncation error. This in particular tests the constraint-damping capabilities of the code. We show the behavior of the perturbed variables in Fig.~\ref{fig:DiabolicalVars}. We see that the solution remains at roundoff level. There is linear growth in $\Delta g_{ab}$, but the level of this growth decreases towards zero with increasing resolution.

\onecolumngrid\newpage\twocolumngrid

\bibliography{dCS_paper}
\end{document}